**World Scientific**
www.worldscientific.com

# Radio Antenna Design for Sky-Averaged 21 cm Cosmology Experiments: The REACH Case

J. Cumner[1,21], E. de Lera Acedo[1,2], D. I. L. de Villiers[3], D. Anstey[1], C. I. Kolitsidas[4], B. Gurdon[1],
N. Fagnoni[1], P. Alexander[1], G. Bernardi[5,6,7], H. T. J. Bevins[1], S. Carey[1], J. Cavillot[8], R. Chiello[9],
C. Craeye[8], W. Croukamp[3], J. A. Ely[1], A. Fialkov[2,10], T. Gessey-Jones[1], Q. Gueuning[1],
W. Handley[1,2], R. Hills[1], A. T. Josaitis[1], G. Kulkarni[11], A. Magro[12], R. Maiolino[1,2],
P. D. Meerburg[13], S. Mittal[11], J. R. Pritchard[14], E. Puchwein[15], N. Razavi-Ghods[1],
I. L. V. Roque[1], A. Saxena[13], K. H. Scheutwinkel[1], E. Shen[1], P. H. Sims[16,17],
O. Smirnov[6,7], M. Spinelli[18,19,20] and K. Zarb-Adami[9,12]

[1]Cavendish Astrophysics, University of Cambridge, Cambridge, UK

[2]Kavli Institute for Cosmology in Cambridge
University of Cambridge, Cambridge, UK

[3]Department of Electrical and Electronic Engineering
Stellenbosch University, Stellenbosch, South Africa

[4]Ericsson AB, Torshamnsgatan 21, 164 40 Kista, Sweden

[5]INAF — Istituto di Radio Astronomia
via Gobetti 101, 40129 Bologna, Italy

[6]Department of Physics and Electronics
Rhodes University, P.O. Box 94, Grahamstown 6140, South Africa

[7]South African Radio Astronomy Observatory
Black River Park, 2 Fir Street, Observatory
Cape Town 7925, South Africa

[8]Antenna Group, Université Catholique de Louvain
Ottignies-Louvain-la-Neuve, Belgium

[9]Physics Department, University of Oxford, Oxford, UK

[10]Institute of Astronomy, University of Cambridge, Cambridge, UK

[11]Tata Institute of Fundamental Research
Homi Bhabha Road, Mumbai 400005, India

[12]Institute of Space Sciences and Astronomy
University of Malta, Msida, Malta

[13]Faculty of Science and Engineering
University of Groningen, Groningen, The Netherlands

[14]Department of Physics, Imperial College London, London, UK

[15]Leibniz Institute for Astrophysics, Potsdam, Germany

[16]McGill Space Institute, McGill University, Montréal, Canada

[17]Department of Physics, McGill University, Montréal, Canada

[18]INAF — Osservatorio Astronomico di Trieste
Via G.B. Tiepolo 11, I-34143 Trieste, Italy

[19]IFPU — Institute for Fundamental Physics of the Universe
Via Beirut 2, 34014 Trieste, Italy

[20]Department of Physics and Astronomy
University of the Western Cape
Robert Sobukwe Road, Bellville 7535, South Africa

[21]jmc227@cam.ac.uk

Following the reported detection of an absorption profile associated with the 21 cm sky-averaged signal from the Cosmic Dawn by the EDGES experiment in 2018, a number of experiments have been set up to verify this result. This paper discusses the design process used for global 21 cm experiments, focusing specifically on the Radio Experiment for the Analysis of Cosmic Hydrogen (REACH). This experiment will seek to understand and compensate for systematic errors present using detailed modeling and characterization of the instrumentation. Detailed quantitative figures of merit and numerical modeling are used to assist the design process of the REACH dipole antenna (one of the two antenna designs for REACH Phase I). This design process produced a 2.5:1 frequency bandwidth dipole. The aim of this design was to balance spectral smoothness and low impedance reflections with the ability to describe and understand the antenna response to the sky signal to inform the critically important calibration during observation and data analysis.

## 1. Introduction

The formation of the first stars in the universe, and associated events, leads to a small but distinct global cosmological 21-cm signature (Furlanetto *et al.*, 2006). The key quantity in deciding this is the spin temperature (quantifying the relative population of hyperfine levels of a neutral hydrogen atom), which in turn is determined by the cosmic microwave background (CMB) temperature, gas temperature and Wouthuysen–Field effect (Wouthuysen, 1952; Field, 1958). The latter is caused by the Lyman-$\alpha$ photons produced by the first stars at Cosmic Dawn (Mittal & Kulkarni, 2021).

The main feature of this signal is a trough visible against the CMB, Fig. 1. Theoretically the depth of this trough is expected to be between 50 and 250 mK considering conventional cooling mechanisms (Fialkov & Barkana, 2019). The detection by EDGES instead placed the depth at 500 mK requiring more exotic cooling mechanisms than previously expected (Bowman *et al.*, 2018).

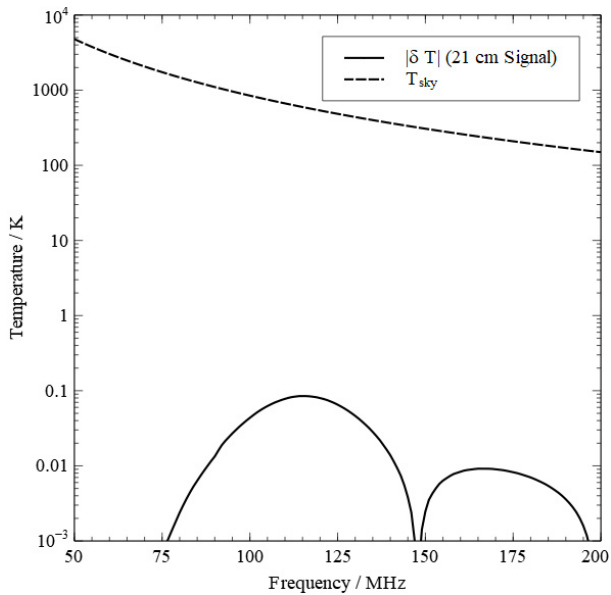

Fig. 1. The magnitude of the expected sky temperature, modeled as a $-2.5$ power law, against the absolute magnitude of an example global 21 cm signal. There is expected to be at least four orders of magnitude difference between the maximum 21 cm signal and the sky temperature at the frequency band of interest here.

The aim for the Radio Experiment for the Analysis of Cosmic Hydrogen (REACH) experiment (de Lera Acedo, 2019) is to provide a confident detection of the 21 cm hydrogen signal from the Cosmic Dawn and the Epoch of Reionization (EoR). The background spectrum being up to five orders of magnitude larger than the signal itself means that to make a confident detection a detailed understanding of the combined effects between the expected power spectrum of the sky and the beam of the antenna is required. A number of experiments aim to measure the 21 cm signal, notably Experiment to Detect the global EoR Signature (EDGES) which reported a detection of the signal in 2018 (Bowman *et al.*, 2018). Many other single-instrument global 21 cm signal experiments, including EDGES (Bowman *et al.*, 2018), Shaped Antenna measurement of the background RAdio Spectrum (Singh *et al.*, 2018a), Probing Radio Intensity at high-Z from Marion (Philip *et al.*, 2019), Mapper of the IGM Spin Temperature (http://www.physics.mcgill.ca/mist/), and Broadband Instrument for global Hydrogen Reionization Signal (Sokolowski *et al.*, 2015) are also searching to make a detection. These instruments lack the angular resolution required to easily perform sky source calibrations and so require a more detailed understanding of the antenna beam and receiver for the calibration process, as compared to interferometers such as the Square Kilometer Array (low band antenna) (SKALA) (Dewdney *et al.*, 2009), Hydrogen Epoch of Reionization Array (HERA) (DeBoer *et al.*, 2017), Large Aperture Experiment to Detect the Dark Ages (Price *et al.*, 2018), and Low-Frequency Array (van Haarlem *et al.*, 2013).

This paper discusses the use of a quantitative figure of merit-based antenna design method, using physically parametrized computer simulations to produce a polynomial fit to describe the individual and combined figures of merit, allowing the unbiased identification of an optimal design from a range of initial designs.

Section 2, *Design Considerations*, discusses the merits of differing antenna styles and their interactions with sky maps used for data analysis. Section 3, *Antenna Design Considerations and Figures of Merit*, details the figures of merit used for the design of the REACH dipole antenna divided into generic antenna figures of merit and those specifically relevant to global 21 cm experiments. An overview of the quantitative figure of merit-based antenna refinement process is given in Sec. 4, *Design Process*. This design process is then implemented in Sec. 5, *Example Case: REACH Dipole* to produce a dipole for the REACH global 21 cm detection experiment.

## 2. Design Considerations

Sky-averaged 21 cm cosmology experiments are based on detecting very weak spectral (across frequency, $\nu$) perturbations of the power measured by a single antenna radiometer, which integrates the sky signal across angle in the sky ($\Omega$) and time ($t$) during an observation. For simplicity, this paper will refer to equivalent temperature quantities instead of power.

The antenna temperature, as a function of time and temperature for a single antenna, is given as

$$T_{ant}(\nu, t) = \frac{1}{4\pi} \int_\Omega D(\nu, \Omega) \eta(\nu) (T_{sky}(t, \nu, \Omega) + T_{21}(\nu)) d\Omega \quad (1)$$

comprising of the antenna beams directivity, $D(\nu, \Omega)$ weighted by the radiation efficiency of the antenna, $\eta(\nu)$, integrated over all angles with the background sky temperature, $T_{sky}(t, \nu, \Omega)$, and the global 21 cm signal, $T_{21}(\nu)$. Typically, the time integral of this antenna temperature is considered for measurements as

$$T_A = \int_t T_{ant} dt. \quad (2)$$

The antenna temperature can also be modified to include impedance reflection and noise effects to give the overall system temperature (Fig. 2)

$$T_{sys}(\nu) = \int_t (T_{ant}(1 - |\Gamma_A(\nu)|^2) G_{RX}(t, \nu) + N(t, \nu, \Gamma_A)) dt. \quad (3)$$

Here, the temperature power is then amplified by the receiver gain, $G_{RX}$, weighted by the antennas' reflection coefficient, $(1 - |\Gamma_A(\nu)|^2) G_{RX}(t, \nu)$. A post-amplification system noise term is added, $N(t, \nu, \Gamma_A)$. The angular scale of fluctuations in $T_{21}$ is expected to be of the order 1 degree on the sky, compared to the tens of degrees of the antenna beam, and so $T_{21}$ can be approximated as having no angular dependence (Liu *et al.*, 2013). It is worth noting that the directivity and the antenna reflection coefficient terms are considered here to be time-independent. In reality, they are not, but the assumption made here is that changes in these parameters will mostly occur at time scales much longer than the time that observation data is integrated for. This assumption should be carefully considered for specific cases (e.g. observations from locations with rapidly changing atmospheric conditions). As in field, calibration methods are used to compensate for the antenna impedance match and excess noise terms, the remainder of this paper will typically use $T_A$ rather than $T_{sys}$ for calculations.

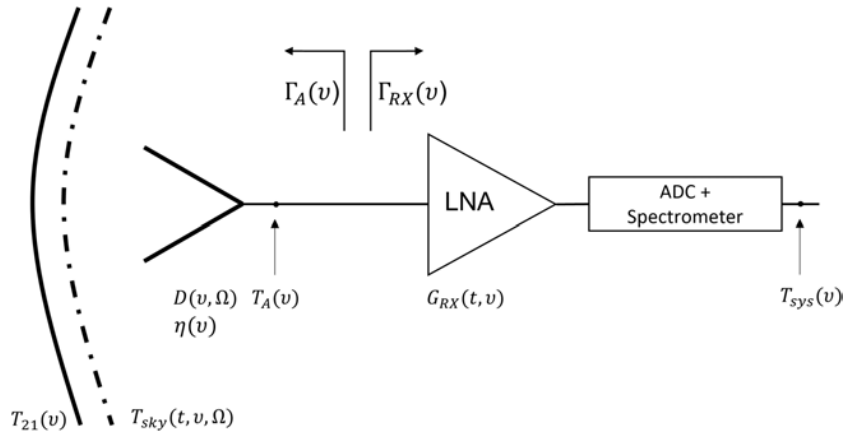

Fig. 2. Diagram of a basic sky-averaged radiometer highlighting the main components of (3). The noise term, $N(t, \nu)$, is to be taken in this paper as the result of a combination of noise components generated at different stages of the chain depicted by this diagram, and before the time integration takes place.

The remainder of this section discusses some initial design considerations required for a global 21 cm antenna. We detail the consideration of the choice between narrow and wide beam antennas, and the uncertainty present in the sky model used. The following section will discuss a selection of figures of merit to quantify antenna characteristics.

## 2.1. *Calibration versus spectral smoothness*

There are three fundamental options in sky-averaged 21 cm experiments. Either the telescope is designed to introduce small enough spectral perturbations in the final system temperature, or later stage calibration implemented to compensate for these effects, or a combination of both. In this paper, the assumption is made that the receiver effects can be calibrated using a combination of measurements and receiver models (Rogers & Bowman, 2012), this assumption is not made for the antenna structure itself. Accurate measurements of a single antenna operating at these wavelengths (1.5–6 m) are extremely difficult and unlikely to be sufficient on their own to calibrate receiver effects. In order to achieve a large enough bandwidth (at the very least 2:1) to be able to leverage on the spectral differences between the 21 cm signal and the foregrounds, it becomes an almost impossible task to design an antenna that will not introduce any (or small enough) spectral perturbations. Thus, it shall be argued throughout this paper that designs for a good radio antenna for this type of experiments need to consider both the ability to calibrate the antenna and inherent spectral smoothness.

As discussed in the following subsection, the ultimate effect of the antenna design in the system temperature does not depend only on the antenna alone, but as indicated by (1), on how the antenna beam couples to the sky signal across time, frequency, and angle. Therefore, it is required to take into account the lack of knowledge of the foreground component of the sky when designing a radio antenna for this application.

Typically, first generation global 21 cm experiments, such as SARAS2 (Singh *et al.*, 2018a), have aimed to construct a global 21 cm antenna with an achromatic beam. REACH will compensate for the beam chromaticity using a Bayesian analysis jointly fitting the cosmological signal, with a modeled sky temperature and an antenna beam model (Anstey *et al.*, 2021b). This means that rather than directly designing the antennas for an achromatic beam, the simplicity of modeling the antenna and its beam is a critical factor in the design of the REACH antennas. Another advantage of this consideration in the data analysis is allowing a broader frequency range to be considered than typically used with low chromaticity antennas, with EDGES-low observing over 50–100 MHz. REACH in contrast will aim to operate a similar dipole over 50–135 MHz. This larger observation range allows for the easier separation of foreground and 21 cm signal and is required by the large variety of possible shapes and locations for the 21 cm signal (Fialkov & Barkana, 2019). Using a single continuous range of frequencies for the antenna increases the likelihood of the entire signal being contained within the observed band so increasing the chance of a confident detection. For the REACH system a conical log-spiral antenna will also be used over a 50–170 MHz band with a less chromatic beam. The increased physical complexity of the conical log-spiral antenna means greater difficulty with modeling and calibration than a dipole antenna. Both antennas will observe from a site in the Karoo Radio Astronomy Reserve near the town of Carnarvon in the Great Karoo semi-desert in South Africa.

## 2.2. *Wide beams versus narrow beams*

For the detection of the 21 cm signal, the approach of taking a sky-averaged measurement to observe a "global" signal can be adopted. This application can lend itself well to a single wide beam antenna, so the averaging is done by the instrument itself. This approach allows for a smaller instrument and reduction in complexity of the required physical equipment. While the physical equipment is simpler the understanding of the system temperature (3), is made more complex if larger portions of the sky are included in the observation, see Fig. 3. There is also the possibility that extremely wide beams, above $60°$, will start to view the horizon. This introduces additional ionospheric effects, radio-frequency interference and increased impact from the ground and soil around the antenna.

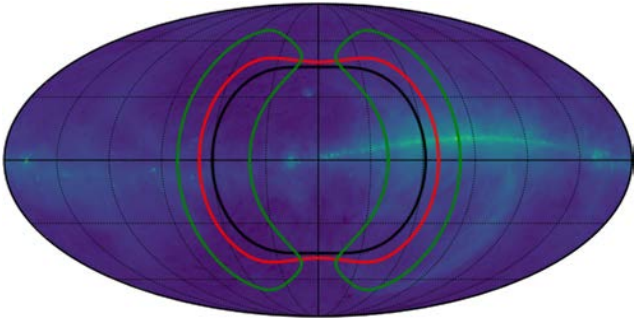

Fig. 3. (Color online) An overlay of the beams of a sample dipole at 50 MHz (black), 100 MHz (red), and 150 MHz (green) with the contour showing the 3 dB gain contour. Here, it is seen that even a relatively narrow beam covers a large section of the sky. This area of the sky will also at times include the galaxy resulting in a complex interaction between beam and sky. A beam smooth over frequency and space will allow for an easier calculation of this interaction.

To understand better the effect of the antenna beamwidth on the detection of the global EoR signal five different antennas are evaluated, in particular, the theoretical model of a horizontal dipole over an infinite ground plane (Balanis, 2005), a simulated realistic bow-tie like dipole (Yang *et al.*, 2016), the SKALA1 antenna that is a log-periodic antenna (de Lera Acedo *et al.*, 2015), the HERA antenna which is a reflector antenna (DeBoer *et al.*, 2017), and a capacitively loaded antenna array (CLDA) which is a simplified design based on the previous work (Kolitsidas *et al.*, 2014). The selected antennas are representative candidates for the EoR signal with two of these, the HERA and SKALA, to be directly involved in instruments that are being deployed to detect the fluctuations of the EoR signal. Bow-tie like dipoles have also been used in the Murchison Widefield Array (MWA), an SKA precursor instrument (Neben *et al.*, 2016). The selected antennas are classified into two categories based on their respective capabilities of angular resolution. For this section focus is placed on the chromatic beam effects of each antenna. The fundamental descriptive properties of the antennas are summarized in Table 1.

An advantage of a high-resolution antenna would be the ability to avoid the complex modeling of the foreground sky signals required with a wider beam antenna, and so increasing the likelihood of a detection (Liu *et al.*, 2013). High-resolution antennas are achieved through either an array or large antenna structure, which means that the beam will have a higher chromaticity than a lower resolution antenna due to additional structures introduced towards the horizon. Array structures also introduce mutual coupling effects in addition to more complex calibration requirements compared to a single small low-resolution antenna. The differing advantages and drawbacks for high- and low-resolution antennas motivate classification into two different groups for a study of their respective effects on system temperature.

Table 1. Summary of the properties of the evaluated antennas. The indicated properties are the min and max directivity $D$ on a dB scale, the voltage standing wave ratio (VSWR), and the angular resolution.

| Antenna | Antenna property | | | |
|---|---|---|---|---|
| | Min $D$ | Max $D$ | VSWR | Ang. res. |
| Ideal dip. | 3 | 7 | — | Low |
| Bow-tie dip. | 3 | 8 | 2.5 | Low |
| SKALA | 3 | 9 | 2.6 | Low |
| HERA | 6.5 | 29 | 2.2 | High |
| CLDA | 7 | 30 | 2.2 | High |

The realistic antennas have been simulated with the time domain finite integral method solver from Computer Simulation Technology — Microwave Studio,[a] with the exception of the ideal dipole that is analytically described. Standard sensitivity testing was performed to verify that during the simulations no numerical resonances occur and a smooth convergence is obtained. The antennas are assumed to be located in the South African Karoo Astronomy Reserve, with coordinates approximately $30.71°$ S and $21.45°$ E. In all simulations a infinite perfect electric conductor is used as a ground plane to remove edge effect reflections and so consider only the chromaticity from the antenna structure. For a finite ground plane the edge reflections will induce a sinusoidal type ripple into the antenna beam, which for a more detailed analysis should be added to the infinite ground plane simulation, or finite ground plane simulations used.

---

[a]https://www.3ds.com/products-services/simulia/products/cst-studio-suite/.

### 2.2.1. *Low angular resolution antennas*

Global 21 cm experiments have, so far, been utilizing primarily wide beam antennas such as dipoles. This results in sampling the radio sky with low angular resolution. Defining low angular resolution antennas in this study as antennas that have half-power beamwidth (HPBW) $> 20°$ and for this analysis this includes the ideal dipole, the simulated dipole and the SKALA.

The directivity of the simple horizontal dipole over an infinite ground plane can be described by the analytical formula based on Balanis (2005):

$$D(\theta, \phi) = 4\sin^2(kh\cos\theta)(1 - \sin^2\theta\sin^2\phi)/R, \quad (4)$$

$$R = \left[\frac{2}{3} - \frac{\sin(2kh)}{2kh} - \frac{\cos(2kh)}{(2kh)^2} + \frac{\sin(2kh)}{(2kh)^3}\right], \quad (5)$$

where $\theta, \phi$ are the zenith and azimuthal angles, $\lambda$ is the wavelength, $h$ is the height above the ground plane, and $k = 2\pi/\lambda$ is the wave number. A schematic representation of the dipole is depicted in Fig. 4 (left). The beam chromaticity is a slowly varying function based on the frequency-dependent ratio $kh$. It is worth noting that an approximation is not taken for (5) since it will not hold for the entire frequency band over which the dipole is evaluated. The parameter $h$ is selected within the range $\lambda_{h\nu}/3 < h < \lambda_{h\nu}/2$, where $h\nu$ denotes the high end of the frequency band. This constraint ensures that only one lobe is achieved within this frequency range minimizing the antenna chromaticity. The obtained beam pattern is illustrated in Fig. 5 (left) at 150 MHz projected into the celestial sphere in equatorial coordinates. This model is included in the present study as a reference case for slowly varying chromatic beams compared to a realistic antenna model.

The second model is a realistic bow-tie like dipole antenna. Bow-tie dipoles utilize tapering to improve the bandwidth performance over the standard dipole. This antenna will be benchmarked directly with the ideal dipole to observe the differences in the performance as global EoR probes. For a comparison at the same frequencies, the same dimensions are used with the bow-tie dipole and ideal dipole and only optimized the width and the tapering. As previously discussed, differences in the distance over the ground plane will result in different chromaticity in the upper half of the frequency band. The schematic of the designed dipole antenna is depicted in Fig. 4 (center). The resulting beam at 150 MHz, as projected into the celestial sphere in equatorial coordinates, is depicted in Fig. 5 (center).

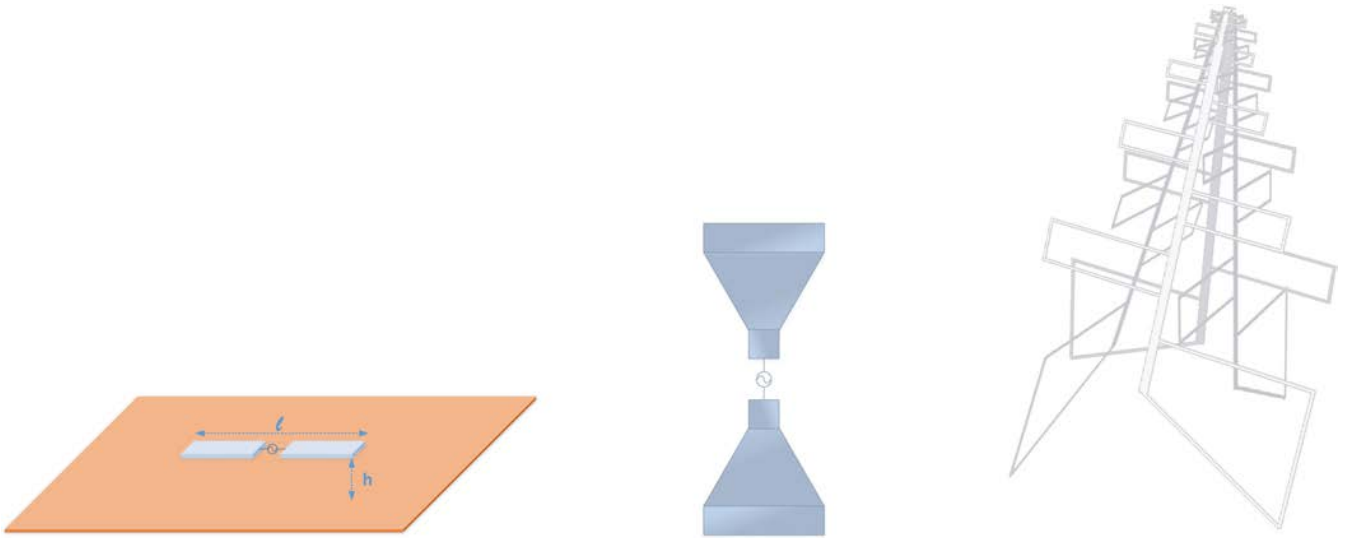

Fig. 4. (Color online) Antenna schematics for ideal dipole (left) simulated dipole (center), and SKALA (right).

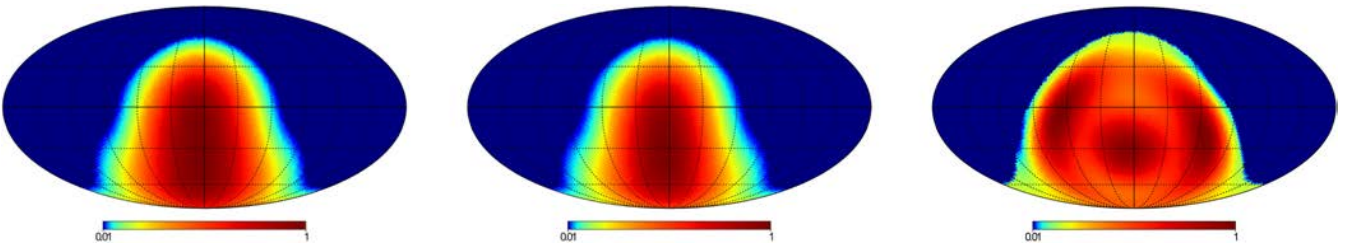

Fig. 5. (Color online) Normalized antenna directivity pattern at 150 MHz projected into the celestial sphere in equatorial coordinates for the (left) ideal dipole over a ground plane (center) dipole simulated in CST, and (right) SKALA.

The final antenna evaluated in the subclass of the low angular resolution antennas is a single SKALA antenna. This is a variation of a dual-polarized log-periodic dipole antenna array. This type of antenna is potentially able to provide an appropriate large bandwidth to probe the EoR with very good inherent impedance matching. An illustration of the SKALA is depicted in Fig. 4 (right). The benefit of the SKALA antenna is that part of the reflector is integrated in the antenna itself for most of the frequency band thereby minimizing the soil and ground effects. The SKALA antenna has not been designed to be used as a standalone antenna but as part of phased array and furthermore it has been designed to cover a large frequency band (7:1), and to provide high sensitivity, which inherently introduces some level of chromaticity on the antenna performance. An analysis of the chromaticity requirements for SKA and the SKALA design can be found in de Lera Acedo *et al.* (2017). This effect is clearly visible with the sidelobes appearing in the radiation pattern as depicted in Fig. 5 (right). The SKALA choice can be considered as a more advanced but realistic implementation of the ideal dipole as better impedance matching can be achieved.

The waterfall diagrams for the low angular resolution antennas are illustrated in Fig. 6 noting the different scales for each antenna. There is approximately an 800 K difference between the ideal dipole model and the simulated bow-tie dipole at the peak temperature. The antenna temperature of the SKALA is approximately 2000 K higher due to the enhanced directivity.

### 2.2.2. *High angular resolution antennas*

An alternative to the wide beam antennas described in the previous section are the high gain antennas with a narrower beamwidth. These antennas can offer higher angular resolution and have not yet been explored extensively in the literature for this type of experiments. Antennas that naturally fall into this category are reflector antennas, lens antennas, large horns, and antenna arrays. Lens antennas are typically based on dielectrics and this will add to the thermal noise of the antenna hence will not be explored further here. The main benefit of a well-designed highly directive antenna is that the field of view can potentially be aligned with cold regions of the sky.

As a first model to evaluate from the class of high angular resolution antennas comes from the reflector type antennas and specifically a single HERA reflector antenna (DeBoer *et al.*, 2017). Reflector antennas have a long and successful history in radio astronomy. The HERA antenna is a reflector with no mechanically moving parts and the beam oriented always to the antenna zenith. The feeder for the HERA dish in this analysis is the dipole initially devoted to Precision Array for Probing the Epoch of Reionization (PAPER; Parsons *et al.*, 2010). The design is depicted in Fig. 7 (left). This design has the benefit that the majority of the collected sky signal is reflected from a metallic mesh reflector to the focal point. Therefore, the effects of the surroundings and the ground are minimized. In addition, HERA's high gain narrow beamwidth with low sidelobes will provide better receiver sensitivity. The simulated normalized beam as projected into the celestial sphere at 150 MHz is illustrated in Fig. 8 (left).

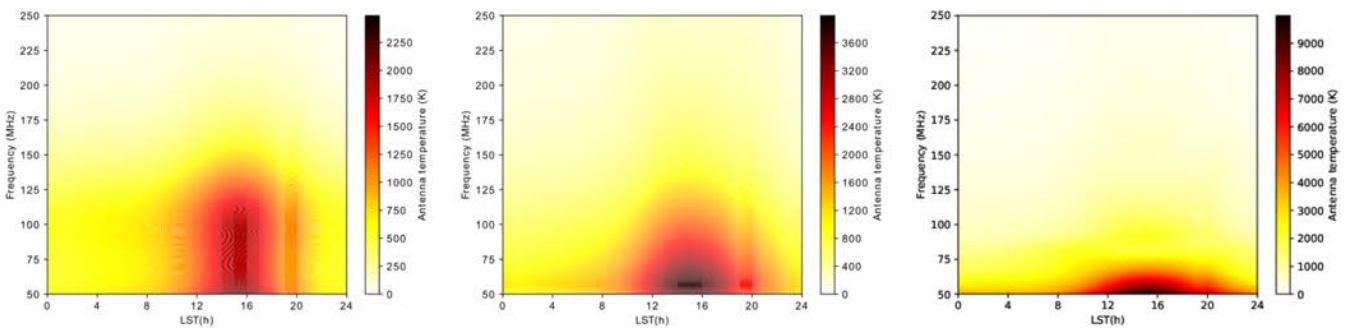

Fig. 6. (Color online) Waterfall diagram of the simulated antenna temperatures for a single day (July 22, 2018) located in the Karoo of (left) ideal dipole (center) simulated dipole, and (right) SKALA. Showing the antenna temperature over frequency and time over 1 day, the galaxy rising over the course of the day causing the rise in antenna temperature seen over time. Note the different scales for the antenna temperature.

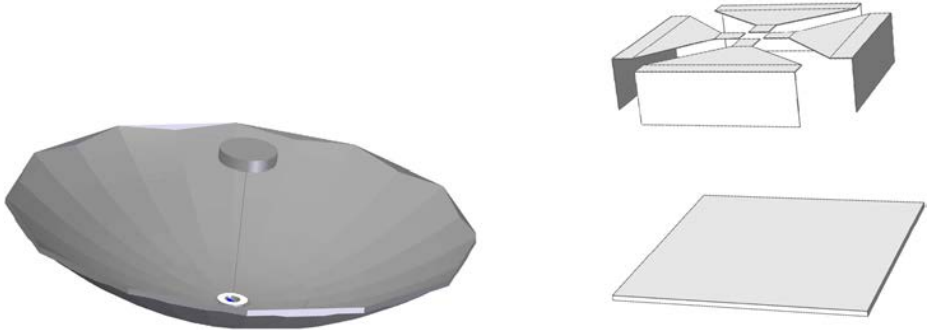

Fig. 7. Illustration of the antenna design of (left) HERA and (right) the unit cell of the CLDA.

Finally, a dense fully populated capacitively loaded dipole array is evaluated as an alternative solution. In Fig. 7 (right), the designed array's unit cell is illustrated. The array is dual polarized and consists of $25 \times 25$ elements per polarization totaling an overall 1250 element ports. The corresponding normalized beam as projected into the celestial sphere at 150 MHz is illustrated in Fig. 8 (right) where the $\sin x/x$ pattern of the uniform aperture illumination is clearly visible. The proposed CLDA can provide similar bandwidth to the SKALA and HERA but in addition it can add certain flexibility, i.e. the beam could scan a cold region of the sky. This reduces the integration time and improves the required stability from the receiver electronics. One could also obtain a reduced chromaticity of the array beam by employing wavelength scale sub-arraying to have an almost constant beamwidth across the frequency band. The CLDA was designed to have almost the same physical area as the HERA dish. The obvious disadvantage of the CLDA is the required control points, hence it is an expensive solution. Another possibility that can be considered is a mechanically steerable reflector or a reflector with a phased array feed.

In Fig. 9, the waterfall diagrams for the antenna temperature are depicted. Observe that when the galactic center is at the antenna zenith the CLDA has double antenna temperature compared to the HERA antenna. On the contrary, during the putative observation time (0:00–4:00 LST) the CLDA and the HERA antenna temperature is similar and approximately 1500 K. It is also observed that in the low frequencies ($\nu \in [50-75]$ MHz) the HERA antenna has higher antenna temperatures than the CLDA.

Based on the waterfall diagrams of all the antennas it is chosen to simulate virtual antenna temperatures between 0:00 and 4:00 LST for every 15 min for an 8 day span. This data is then used for all calculations referred to in the following analysis. The 8 day span is selected to simulate the conditions of a realistic experiment and further smoothing of the antenna temperature.

### 2.2.3. *Foreground modeling for different antenna types*

At the frequencies where the global 21 cm signal is expected (50–200 MHz) the radio sky is dominated by synchrotron emission (Jelić *et al.*, 2010). The synchrotron emission has been proven to follow a simple power law relation (Jelić *et al.*, 2010), and the scaling factor is usually referred to as the spectral index $\beta$. Equation (6) reads the frequency scaling relation of the radio sky referenced to the entire sky survey of the 408 MHz Haslam map (Haslam *et al.*, 1982). The CMB is not scaled and the Planck data (Adam *et al.*, 2015), is used for the CMB performing the operation at every pixel. The scaled map for each frequency $\nu$ is given as

$$T_{\text{sky}}(\nu, \mathbf{s}) = (T_{\text{Sky Model}}(\mathbf{s}) - T_{\text{CMB}}) \left(\frac{\nu}{\nu_{\text{sky}}}\right)^{-\beta} + T_{\text{CMB}}. \tag{6}$$

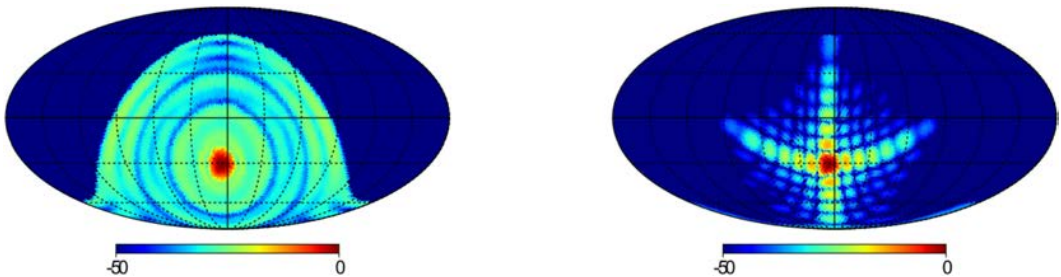

Fig. 8. (Color online) Normalized antenna directivity pattern in dB scale at 150 MHz for (left) simulated HERA and (right) simulated CLDA with uniform aperture illumination.

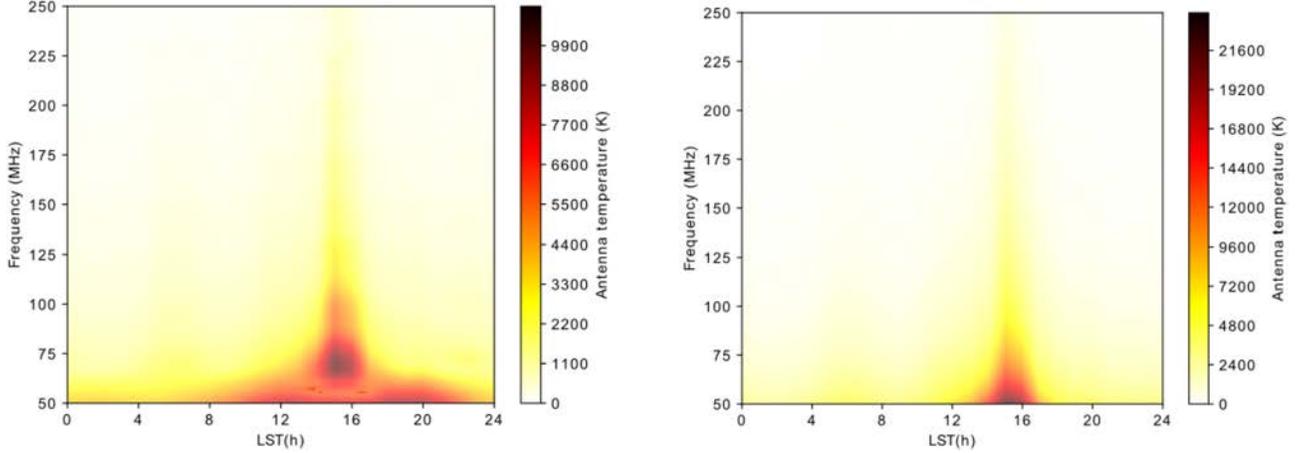

Fig. 9. (Color online) Waterfall diagram of the simulated antenna temperatures of (left) HERA and (right) CLDA.

In order to do an initial evaluation of the sky-antenna spatial-spectral chromaticity in the context of analyzing antennas with different beamwidths (wide or narrow) and its impact on the calibration of foregrounds for global 21 cm experiments, here two different sky models are used. The first sky model is the unfiltered — NF version of the Haslam map (Haslam *et al.*, 1982) with a constant spectral index $\beta = 2.5$, as has been extensively used in the literature (Matteo *et al.*, 2004; Rogers & Bowman, 2008; Bowman *et al.*, 2009). The second sky model is the widely used destriped desourced (DSDS) updated Haslam map from Remazeilles *et al.* (2015). To account for the angular dependency a similar technique to Bernardi *et al.* (2015) and Mozdzen *et al.* (2017) is adopted, where the spectral index is extracted according to equation:

$$\beta = - \left[ \ln \left( \frac{T_{\text{sky}\nu_{\beta}} - T_{\text{CMB}}}{T_{\text{sky408 MHz}} - T_{\text{CMB}}} \right) \middle/ \ln \left( \frac{\nu_{\beta}}{408} \right) \right], \tag{7}$$

where $T_{\text{sky}\nu_{\beta}}$ refers to the sky as extracted from de Oliveira-Costa *et al.* (2008) whereas the $T_{\text{sky408 MHz}}$ is the DSDS Haslam map. The result of this ratio (7) is depicted in Fig. 10 in galactic coordinates, where it is observed that the spectral index is not uniform across the entire sky when calculated pixel by pixel. In this work, the spectral index is calculated only for a single frequency at $\nu_{\beta} = 100$ MHz. The scaled Haslam map at 150 MHz with varying spectral index is illustrated in Fig. 11 in equatorial coordinates. A straightforward extension to this method is to obtain the spectral index for each frequency following the same procedure using (7). This will also account for the frequency variants of the spectral index. From the illustration of the spectral index in Fig. 10, it is expected that the varying spectral index is more suitable for the high angular resolution antennas as a smaller part of the sky is sampled with the antenna beam whereas a constant spectral index can be utilized in the low angular resolution

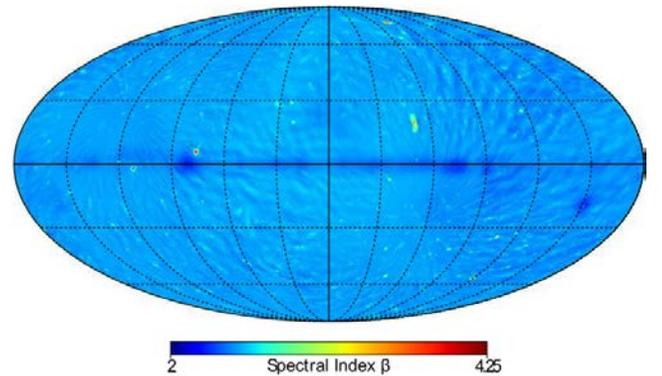

Fig. 10. (Color online) Spectral index $\beta$ in galactic coordinates.

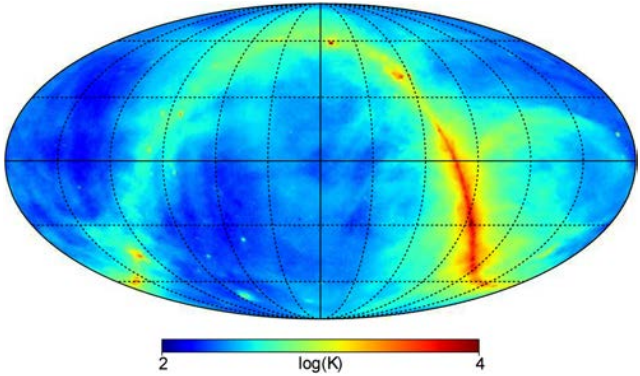

Fig. 11. (Color online) 150 MHz scaled Haslam map in equatorial coordinates using the angular dependent spectral index $\beta$ from Fig. 10.

antennas as the effect will be averaged out to a larger extent.

For an almost achromatic antenna, if it is possible to assume a small contribution of the spectral index distribution (e.g. one single spectral index), a logarithmic polynomial model of the antenna temperature when only foregrounds are observed, as in (8), should be suitable [similar to the one proposed in the works of Pritchard & Loeb (2010) and Wang & Hu (2006)].

$$\log \hat{T}_{\text{ant}}(\nu) = \sum_{n=0}^{m} \alpha_n (\log \nu)^n, \qquad (8)$$

where the $\hat{}$ denotes modeled data.

Then, to assess the impact of the antenna chromaticity on foreground calibration the figure of merit $\text{SFoM}(\nu)_i$ is used, as in the following equation:

$$\text{SFoM}(\nu)_i = \sqrt{|T_{\text{ant}}(\nu)_i - \hat{T}_{\text{ant}}(\nu)|^2},$$
$$i \in \{\text{NF}, \text{DSDS}\}, \qquad (9)$$

where the root mean square (rms) error for each antenna model of the estimated value and the polynomial fit in the entire band for each sky model is evaluated. In the following analysis the two sets $\{\text{NF map}, \beta = 2.5\}$ and $\{\text{DSDS map}, \beta = \text{Fig. 10}\}$ are used and the corresponding set is indicated by the map choice as a subscript. It is noted that in this exercise the data is integrated for 4 h and that here at present only the foreground calibration is under assessment and not any potential signal loss and thus the order of the polynomial model is kept low.

Also defined is the chromaticity correction, $C(\nu)$:

$$C(t, \nu) = \frac{\int_{\Omega} D(\nu, \Omega) T_{\text{sky}}(t, \nu_0, \nu) d\Omega}{\int_{\Omega} D(\nu_0, \Omega) T_{\text{sky}}(t, \nu_0, \nu) d\Omega}. \qquad (10)$$

This is used as a scaling factor to reduce errors in antenna temperature due to chromaticity in an antenna beam, directivity $D(\nu, \Omega)$, with a sky temperature, $T_{\text{sky}}$, to a reference frequency, $\nu_0$. The selection of an appropriate sky map is necessary for an accurate correction to an antenna temperature. With a consistent sky map this figure can be used to quantify and compare the chromaticity present in various antenna beams.

The ideal dipole and the simulated bow-tie dipole are evaluated and the results are presented in Fig. 12. It is observed that for the ideal dipole the rms error is very small ($\leq 5$ mK) in both sky models evaluated as well as for only fourth-order polynomial fit. The results are illustrated in the zoomed area in Fig. 12. The reason for this behavior is traced to the slowly varying function that represents the ideal dipole. In contrast, the simulated bow-tie like dipole has a more severe impact on the rms error between the fitted data and the calculated one. The average $\langle \text{SFoM}(\nu)_{\text{DSDS}} \rangle = 17$ mK whereas is about $\langle \text{SFoM}(\nu)_{\text{NF}} \rangle = 36$ mK for sixth-order polynomial fitting. This is an indication of the impact of sky model on the calibration of the instrument. In addition, even though the ideal dipole has very low residuals it is hardly the case for a representative realistic antenna model. This implies that in the design of the global EoR experiment the chromatic antenna behavior should be taken into account.

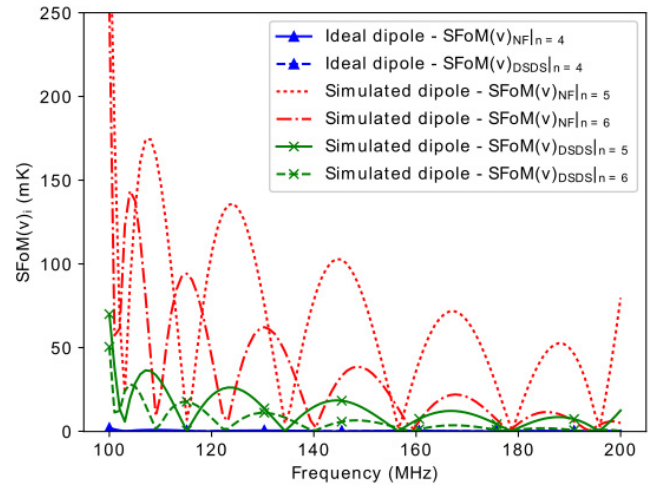

Fig. 12. (Color online) Residuals for the theoretical dipole model and the simulated bow-tie dipole model.

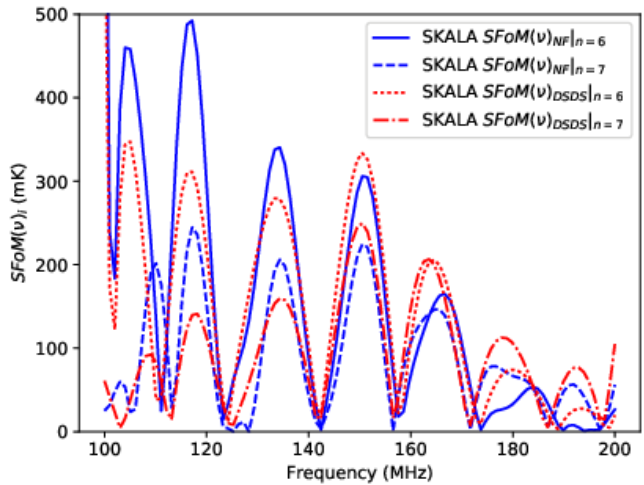

Fig. 13. (Color online) Residual for the SKALA model for the different sky models.

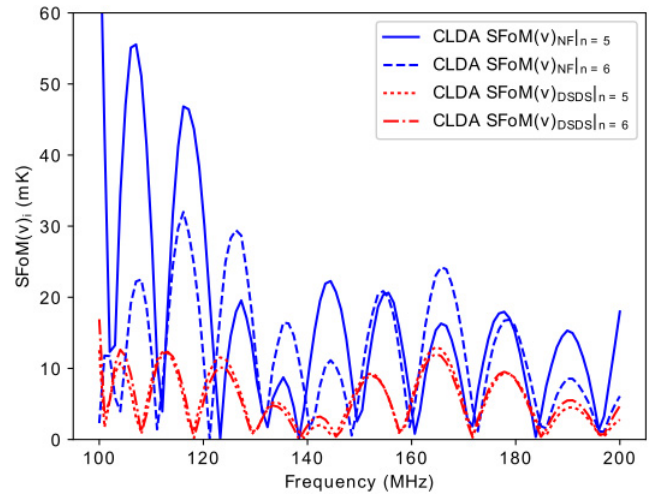

Fig. 15. (Color online) Residual for the CLDA model for the different sky models.

The impact of the antenna chromaticity in the SKALA case is illustrated in Fig. 13, with the data fitted per frequency over the entire interval. The averaged $\langle \mathrm{SFoM}(\nu)_{\mathrm{NF}}\rangle = 104\,\mathrm{mK}$ and $\langle \mathrm{SFoM}(\nu)_{\mathrm{DSDS}}\rangle = 82\,\mathrm{mK}$ for a seventh-order polynomial fitting. These differences stem from the antenna beam chromaticity of the SKALA as well as the limited simulated time interval of observation.

The residuals for the case of the HERA reflector antenna are presented in Fig. 14. Here, it is noted that the scale on the $\mathrm{SFoM}(\nu)_i$ axis is K. Observe that the $\langle \mathrm{SFoM}(\nu)_i\rangle$ becomes $\langle \mathrm{SFoM}(\nu)_{\mathrm{NF}}\rangle = 1.14\,\mathrm{K}$ and $\langle \mathrm{SFoM}(\nu)_{\mathrm{DSDS}}\rangle = 0.5\,\mathrm{K}$ for a seventh-order polynomial fitting. The reason for these high values is that the HERA antenna radiation pattern, despite being sensitive to a smaller region of the sky compared to wide beam antennas, is also fairly chromatic, resulting in a potentially challenging foreground modeling. This chromaticity is induced due to the reflector being electrically small and substantial diffraction effects being present, in addition to standing wave effects between the feed and the dish. The impact of a noisy, with no angular resolution sky model is visible in Fig. 15 for the case of CLDA. In this case, the antenna has been tuned to present a smooth beam across frequency. The average sky FoM becomes $\langle \mathrm{SFoM}(\nu)_{\mathrm{DSDS}}\rangle = 8\,\mathrm{mK}$ for a sixth-order polynomial fitting that is the lowest for all realistic evaluated antenna models in this study.

Following this discussion, it is concluded that a high angular resolution antenna, sensitive to a smaller region of the sky when compared to a wide beam antenna, does not necessarily translate into an easier foreground modeling, and the beam chromaticity seems, in principle, to be a more important factor. It is also important to use a sky model that provides the correct angular resolution. With this in mind a focus will now be placed on wide beam antennas, which are typically also easier to model and cheaper to build, as well as potentially less chromatic than high gain antennas owing to their smaller electrical size and the smaller number of current modes they can support.

## 2.3. Uncertainty in the sky model

A further complication in the calibration process of a radio antenna beam for global experiments is that much of it relies on some knowledge of the foregrounds. Several approaches have been proposed and used to do this, including no correction at all, a correction based on a chromaticity correction factor (10) (Monsalve *et al.*, 2017) or a Bayesian fitting of a parametric foreground model procedure (Anstey *et al.*, 2021b), as used by REACH. It shall be argued in Sec. 3, that the chromaticity correction factor is a good, and straightforwardly calculated, indicator of the chromaticity present in an antenna beam. An improved method for performing this calculation is to construct a pipeline for injection and attempted detection of a signal through the candidate antenna beam [in the case of REACH, this is the Bayesian foreground calibration algorithm described in Anstey *et al.* (2021b)], this however is too time consuming to be done to thousands of antennas. One of the reasons to do this is the lack of good foreground maps at these frequencies, the sensitivity of figures of merit and less flexible beam corrections to these sky maps. In order to illustrate the impact of the uncertainties introduced by the lack of knowledge of the sky, Fig. 16 shows a simulation of the antenna temperature for a simple square dipole. This antenna temperature, $T_{A,\mathrm{corr}}$, has been corrected using the chromaticity correction factor (Monsalve *et al.*, 2017) using a different sky map than the one used to generate the simulated data. This is then compared to an antenna temperature calculated using the 50 MHz beam at all frequencies, $T_{A,\mathrm{noChr}}$. If the chromaticity correction is perfect, the plotted, value $T_{A,\mathrm{corr}} - T_{A,\mathrm{noChr}}$ should be 0, with the same pair of sky maps used for antenna temperature calculation and correction.

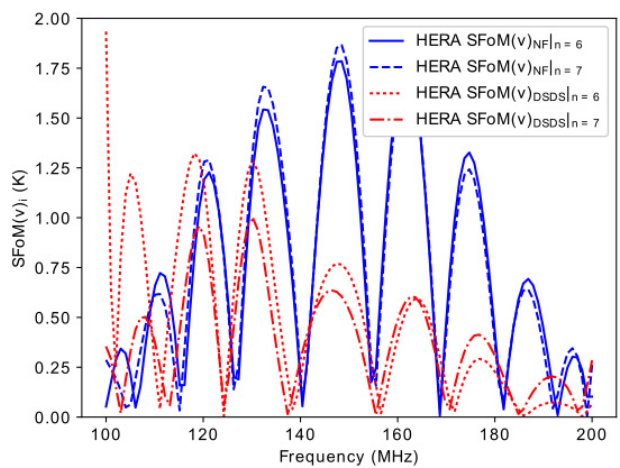

Fig. 14. (Color online) Residual for the HERA model for the different sky models.

In Fig. 16, the sky model used to generate the antenna temperature was the GSM(200) map (de Oliveira-Costa *et al.*, 2008) and the Haslam map (Haslam *et al.*, 1982) was used as the model. This is instructive as the Haslam map scaled with a spectral index of −2.5 was the model used by EDGES in their chromaticity correction (Bowman *et al.*, 2018). The residuals of the chromaticity correction for different locations of the antenna on the Earth surface are shown. All curves show a high degree of non-smooth behavior and residual features of up to −0.6 K in the frequency range 50–100 MHz.

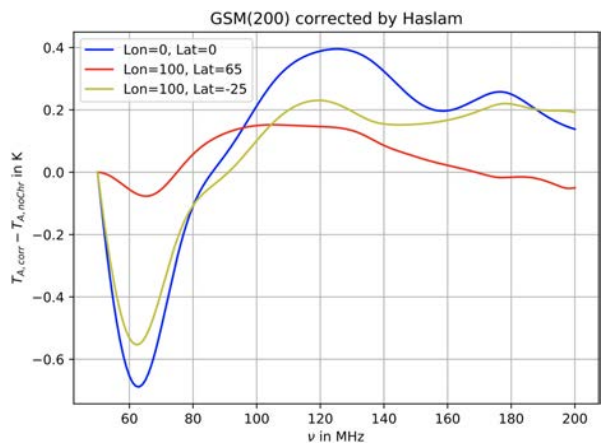

Fig. 16. (Color online) Residuals when correcting the antenna temperature data generated with a GSM sky using a HASLAM sky model.

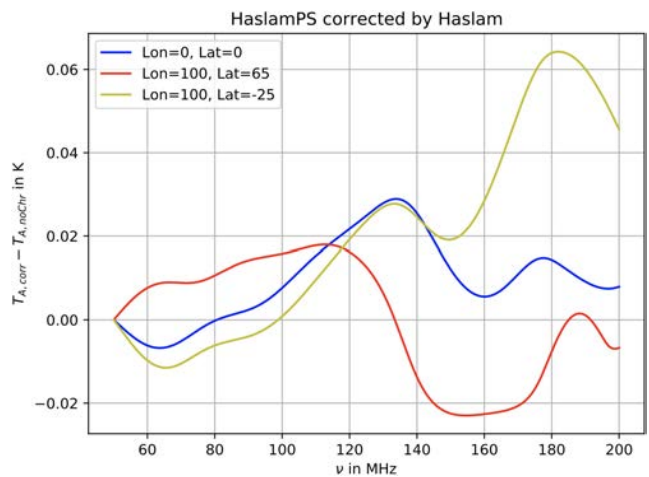

Fig. 17. (Color online) Residuals when correcting the antenna temperature data generated with a HASLAM sky with point sources using a de-sourced HASLAM sky model.

The other example case shown here (see Fig. 17) is on the influence of point sources. For this, the sky is assumed to be the Haslam map with point sources in it while the de-sourced Haslam map is taken as the model. One can see that in this case the residuals are much smaller and only on the order of 10 mK in the range from 50 to 100 MHz.

Note also that similar dependencies are observable when changing the reference frequency for the correction, the integration time, etc., as well as many other factors that would make the sky model different from the real one, such as a per pixel spectral index, polarization, etc. The difficulty of modeling these effects and small uncertainties associated mean that they can be neglected for an initial antenna design. For a confident detection to be made these effects are required to be included in the sky map used for data analysis.

## 3. Antenna Design Considerations and Figures of Merit

For the design process of the REACH antenna a quantitative refinement approach was taken, using a selection of figures of merit to describe and optimize features of candidate antennas. The figures of merit used were a combination of generic antenna figures of merit and those specific to a global 21 cm experiment, aiming to produce low impedance reflections and a low chromaticity beam. The figures of merit were optimized through the refinement of the physical geometry of the antenna and accompanying balun structures. Accompanied by prototyping principal component analysis and mock detection pipeline analysis of promising designs, this forms the recursive process detailed in Fig. 18. Resulting in the join optimization of the antenna input observed from the sky, and the output electrical interaction with the antenna receiver structure.

For the REACH experiment the figures of merit and their goal values were initially chosen based on empirical evidence working with similar antennas and preliminary results from the initial antenna design exploration. These design figures of merit are selected to be representative while quickly calculable and quantifiable, such that a large possible design space can be explored quickly and efficiently with the most promising designs being passed onto more time consuming and rigorous tests such as a mock detection pipeline.

### 3.1. *Generic*

#### 3.1.1. *Antenna impedance figures of merit*

For the REACH system a $50\,\Omega$ coaxial cable is used to connect the antenna structure to the receiver system. The suitability of the one-port scattering parameter, $S_{11}$, which is the impedance reflection coefficient into $50\,\Omega$, is considered through a collection of figures of merit. As the phase of the $S_{11}$ is of low importance for this case the absolute value in dB is considered here.

As calibration error is proportional to the value of $|S_{11}|$ the mean value of this quantity is considered to be minimized as the first figure of merit $\alpha$:

$$\alpha = \langle dB(|S_{11}|)\rangle. \tag{11}$$

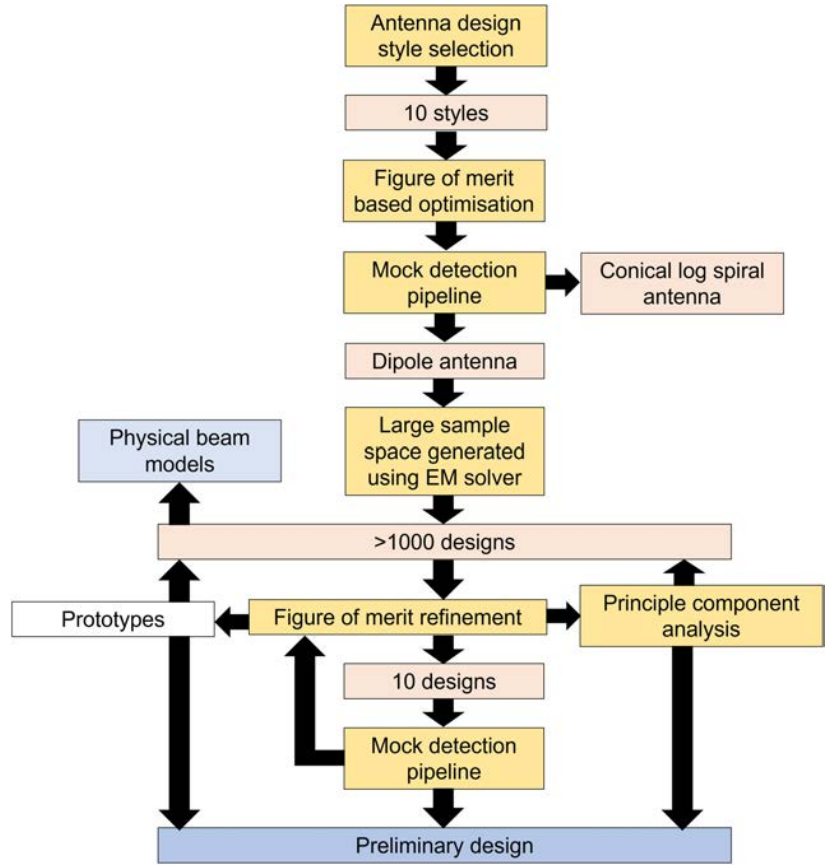

Fig. 18. (Color online) The design process used for the design of the REACH antennas, designed to be unbiased towards antenna design styles, Fig. 23. Starting with the consideration of many styles before refining the chosen design style with an iterative quantitative figure of merit-based refinement.

To assess the possible calibratable bandwidth of the instrument the figure of merit $\beta$ is considered defined as

$$\beta = \nu|_{S_{11}}^{+} = -10\,\text{dB} - \nu|_{S_{11}=-10\,\text{dB}}^{-}, \qquad (12)$$

where $\nu|_{S_{11}=-10\,\text{dB}}^{-}$ and $\nu|_{S_{11}=-10\,\text{dB}}^{+}$ are the minimum and maximum frequencies at which $|S_{11}|$ crosses $-10\,\text{dB}$. If this value is crossed multiple times the lowest pair of frequencies is selected. Theoretical predictions of the 21 cm signal, and the EDGES detection, suggest that this range should cover from at least 60–130 MHz to ensure that the large trough of the signal is included within the operating range of the antenna.

To give an approximation of the shape of the reflection coefficient an integrated area between threshold values and the reflection coefficient magnitude is considered, $\gamma$:

$$\gamma = \left\langle \sum_{i=0}^{6} \int f_i(\nu) d\nu \right\rangle \qquad (13)$$

with

$$f_i(\nu) = \max\{6 + 2i - |\text{dB}(S_{11})|, 0\}. \qquad (14)$$

It is expected that a steeply varying change in the reflection will reduce the accuracy of calibration of the final data and should be avoided, especially in regions where a 21 cm signal is expected to reside. A lower value of $\gamma$ will provide a flatter and lower $S_{11}$ curve (Rogers & Bowman, 2012).

As bandwidth, $\beta$, can be increased through scaling the antenna it is important to act to keep the majority of bandwidth in the relevant section of frequency space, 50–150 MHz. So the fourth figure of merit, to be minimized, $\delta$ is defined as

$$\delta = \nu|_{S_{11}=-10\,\text{dB}}^{-}. \qquad (15)$$

Effective calibration of an antenna can be assisted with an understanding of the phase of $S_{11}$. For a simple dipole the main contribution is from the length of cable connecting the antenna blades and receiver. When physically controllable in this way it is not required to compensate for this effect within the blade design. For other antenna designs, such as a conical log-spiral, the large and comparatively complex balun structure required for feeding and matching the antenna means that the phase of $S_{11}$ would require a figure of merit during the design process.

### 3.1.2. *Engineering and complexity considerations*

In order to allow for an understanding of the antenna as detailed as possible, a simple design is of high importance. For the initial design and final analysis stages this pertains to the computational models used for simulations. The complexity of these models is best quantified in the number of mesh cells required for accurate modeling of the antenna (given antennas of roughly equal size, operating over similar frequencies).

Engineering constraints for the construction and maintenance of the antenna are considered through the use of prototypes constructed in a lab setting. These prototypes are also used to verify simulation results for the antenna, indicating any areas of simulation requiring increased attention.

### 3.2. *Global 21 cm experiment*

#### 3.2.1. *Antenna beam chromaticity figure of merit*

To quantify the chromaticity of the antenna beam a method following from the EDGES chromaticity correction is used (Monsalve *et al.*, 2017). To match to the final data analysis pipeline, the sky map used for calculation of the chromaticity figure of merit is generated using a nine region spectral index sky map, where each region of the sky has a spectral index calculated using the method detailed previously, based on (6) and (7); giving an expected antenna temperature $T_{\text{sky}}$ for the candidate antennas.

This sky map serves as a close approximation of the structures seen within the sky at the frequencies to be observed, allowing for a good representation of the interaction between the beam structure and the structure of the sky for design purposes. For the REACH antenna a broad beam antenna will be used meaning that a large area of the sky will be observed at all times, so averaging out details within the sky.

The chromaticity correction factor (10) provides an approximation of the chromaticity present within an antenna beam. For the figure of merit, $\sigma$ the variance of this time–frequency grid is considered:

$$\sigma = \langle C^2 \rangle - \langle C \rangle^2. \qquad (16)$$

The value of $\sigma$ is sought to be minimized for an optimal antenna, thus giving the smoothest beam response, as during the data analysis it is required to recover the expected sky signal following its interaction with the antenna beam.

### 3.2.2. *Mock detection pipeline*

For the most promising designs a more stringent, and time consuming, analysis is carried out using a mock detection pipeline. For this a selection of possible 21 cm signals are injected into a constructed sky map (Anstey *et al.*, 2021b). These skies are then treated as the input data for a Bayesian process during which the pipeline will attempt to detect the input signals, or lack thereof. This method allows for a more detailed verdict on the beams and using the full interaction between the beams, sky map, signal, and analysis technique. This method is expected to provide the most accurate representation of the effectiveness of an antenna for detecting a global 21 cm signal. However due to the long run time of evaluations, above 7 days, this method is only suitable for a final decision between a small number of the most promising designs identified other figures of merit (Anstey *et al.*, 2021a).

## 4. Design Process

For the quantitative figure of merit design process, an example for REACH shown in Fig. 18, an iterative refinement process was used. This approach was designed to start with as little prejudice as possible towards any particular antenna design, instead focusing on producing an antenna to best synergize with the calibration and analysis stages of the system.

An overview for the design process to accompany the previously detailed figures of merit is shown in Fig. 18. For REACH it was chosen to focus upon wide beam antennas, as any advantage gained from a narrower beam in an array is negated by the increased cost and complexity required in modeling. The first stage of the design process required a collection of different antenna types for an initial rough analysis, narrowing down a dozen designs to one or two. An example of each of the candidate designs was optimized using the figures of merit and then examined using the mock detection pipeline to identify the likelihood of a successful detection being made with each candidate antenna design.

Following the selection of the overall design type the commercial electromagnetic field simulation software CST microwave studio is used to generate a large initial design set. These designs are all processed using numerical figures of merit. This set is then used for a preliminary analysis allowing for expansion into other promising areas of the parameter space.

When a smaller sample set is decided upon the next step is to again apply the mock detection pipeline, which due to time constraints is not usable on large design numbers. This acts as a strong test for the expected interaction of the antenna beams with the sky and 21 cm signal. As the interaction between beam chromaticity and ability to detect a 21 cm signal is less well understood, than the effect of impedance on calibration, this step is used to assist the selection of limits to place on figures of merit.

After a sufficiently capable design is arrived at the refinement loop ceases and a preliminary design is made and deployed to the field to attempt a detection. It is expected that with a first design further refinements will be required due to complications only identified when making observations.

### 4.1. *CST simulation*

The antenna simulations were carried out predominantly using the finite-element-method-based frequency domain solver inbuilt to CST. Typical calculations take an hour of server (16 cores at 2.6 GHz, 387 GB RAM) time to compute, with some variance due to the complexity of design.

For each simulation, the inbuilt CST mesh refinement method, based upon $S_{11}$ convergence, routine was used. Figure 19 shows the convergence for simulation time, mesh cells and $S_{11}$ for a hexagonal dipole. Each simulation was conducted with a minimum of four refinement stages, and is terminated once three consecutive simulations have less than 0.02 fractional difference between $S_{11}$ calculations.

Figure 20 shows the final mesh used for a simulation of an example pair of antenna blades. With the addition of the 10 m radius ground plane, this simulation required $4.25 \times 10^5$ tetrahedral cells. As is shown in the figures the size of these cells varies greatly depending upon the surrounding model structure. Given the presence of quasi-singular currents on the edges of the blades, a finer mesh on the blades contour is desired. With the more detailed areas, such as the edges of the blades, a far more detailed mesh is required to compute. Shown in Fig. 21 are the corresponding reflection coefficients for the different levels of completeness of the antenna, showing the impact of including the balun and ground plane in the simulation, encouraging the optimization of all components at the same time, despite the longer computation times.

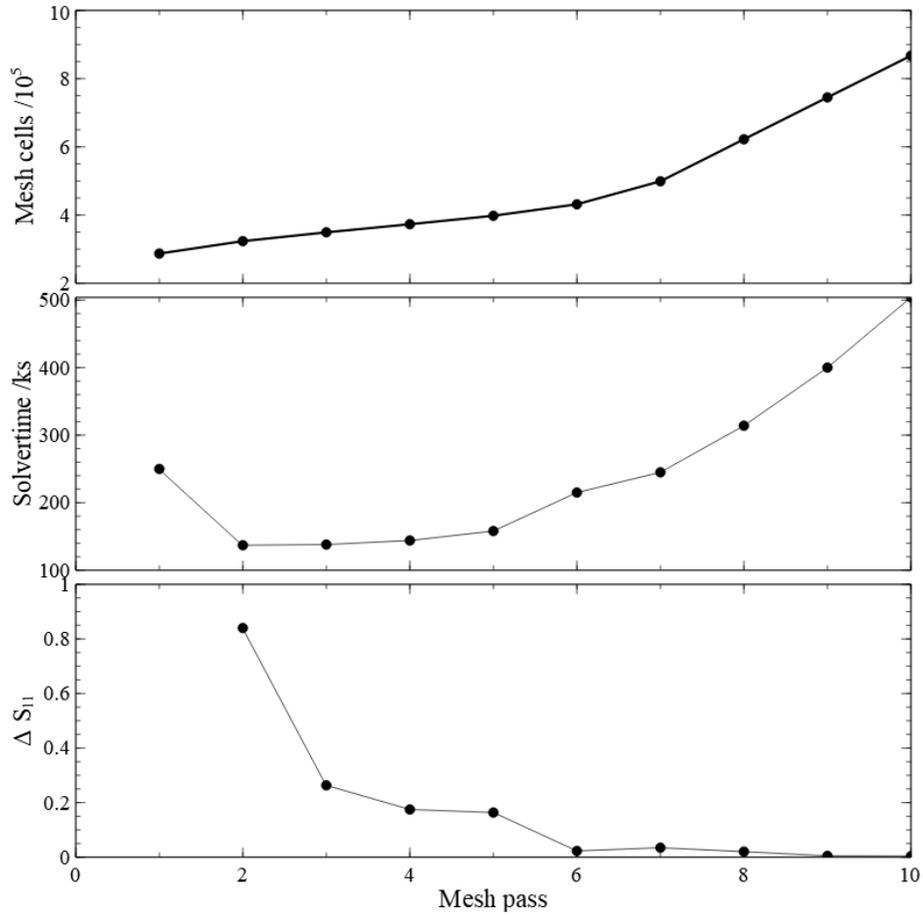

Fig. 19.   The convergence of an example simulation for a dipole, showing the near exponential growth in solver time and mesh cells. Along with the reduction in $\Delta S_{11}$ as the mesh grows finer, for halting the simulation and mesh refinement $\Delta S_{11} < 0.02$ for five consecutive simulations.

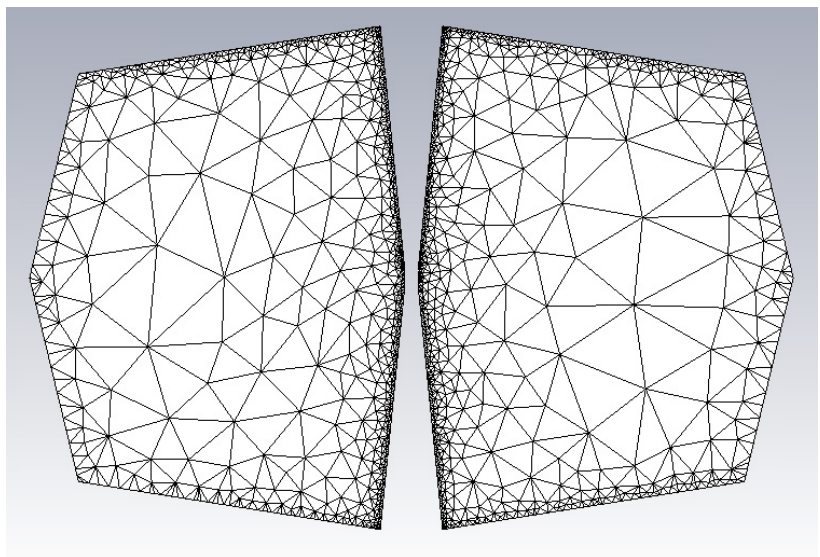

Fig. 20.   Example of a final meshing for antenna blades, showing the grouping of smaller mesh cells at the edges of the metal sheet. Adding smaller scale structure to simulations, has a large impact on the required simulation time.

Table 2. The number of mesh cells required to simulate components of a broad band dipole. The simulation with only the antenna blades forming the smallest and simplest composition therefore requiring the least mesh cells to accurately simulate. The addition of the large ground plane increases the starting number of mesh cells by an order of magnitude. Adding the smaller scale balun structure to the simulation required double the number of mesh cells for the final pass.

| Component | 1st pass mesh cells | Final pass mesh cells |
|---|---|---|
| Blades only | $1.0 \times 10^4$ | $2.5 \times 10^4$ |
| Balun only | $5.0 \times 10^4$ | $1.0 \times 10^5$ |
| Blades and 5 m ground plane | $8.0 \times 10^4$ | $1.5 \times 10^5$ |
| Blades and 10 m ground plane | $2.5 \times 10^5$ | $4.25 \times 10^5$ |
| Full antenna and 10 m ground plane | $2.5 \times 10^5$ | $1.0 \times 10^6$ |

## 4.2. *Figure of merit scaling*

To allow comparison between the various figures of merit scaling is required. Two methods are used for this step, one for an initial exploration of the unknown parameter space and the second is applicable once quantified requirements for the figures of merit are established. The values of $S_{11}$ are taken from each antenna design from which each of the $S_{11}$ figures of merits is calculated.

For initial parameter space exploration a large and sparse initial sampling set is used. These values are then scaled linearly over all samples, with the most preferential result given a 0 value and the least preferential 1. So in this case the scaled figures of merit $(\phi)$ are

$$\phi_i = \frac{F - F_{\min}}{F_{\max} - F_{\min}}. \tag{17}$$

When limits of figures of merit are more accurately understood a sigmoid function can be used for a more efficient and targeted refinement, the function used for the REACH dipole is

$$\Phi_i = \left(1 + \left(\frac{F}{y}\right)^{\frac{\log(9)}{\log(\frac{z}{y})}}\right)^{-1}, \tag{18}$$

where $y$ is the minimum acceptable value of the figure of merit, so maximized gradient of $\Phi_i$, at $\Phi_i(y) = 0.5$, and $z$ is selected such that $\Phi_i(z) = 0.1$. An example of this scaling function with areas of interest is shown in Fig. 22 using the $\sigma$ figure of merit.

For an experiment including multiple figures of merit which are interacting in such a way that improvement of one reduces the suitability of the other it is required to ensure that over optimization of one figure of merit at the expense of another occurs. For this reason a sigmoid function is well suited compared to a linear scaling. The functional shape means that once a satisfactory value is achieved gains in the scaled figure of merit are reduced, allowing for improvement in others.

Over the course of an experiment the goal values should remain flexible as more is learned about the greatest constraints on a possible successful detection. The first choice of $y$ and $z$ should be such that there is some flexibility, and examining the values of a large selection of candidate antennas to check that the whole sigmoid curve is represented serves as a good check that the problem is not over constrained.

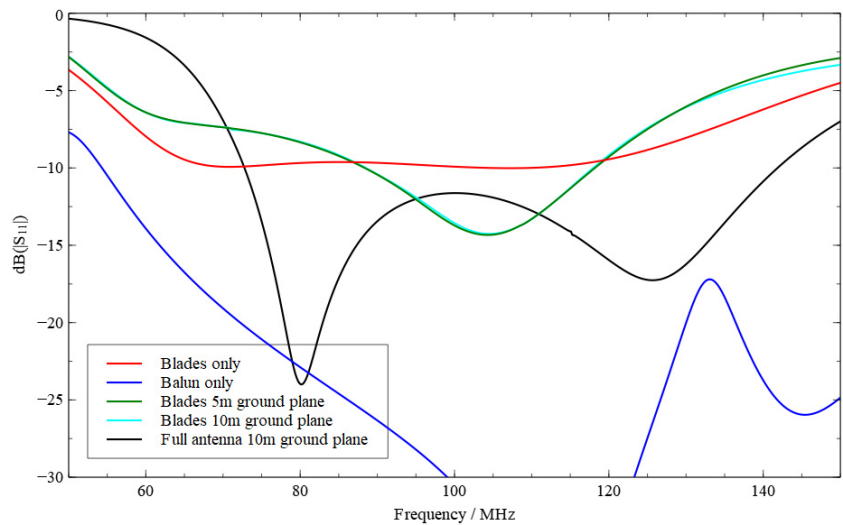

Fig. 21. (Color online) The accompanying scattering parameters for Table 2, demonstrating the need to simulate the entire antenna structure for an accurate measurement of the reflection coefficient.

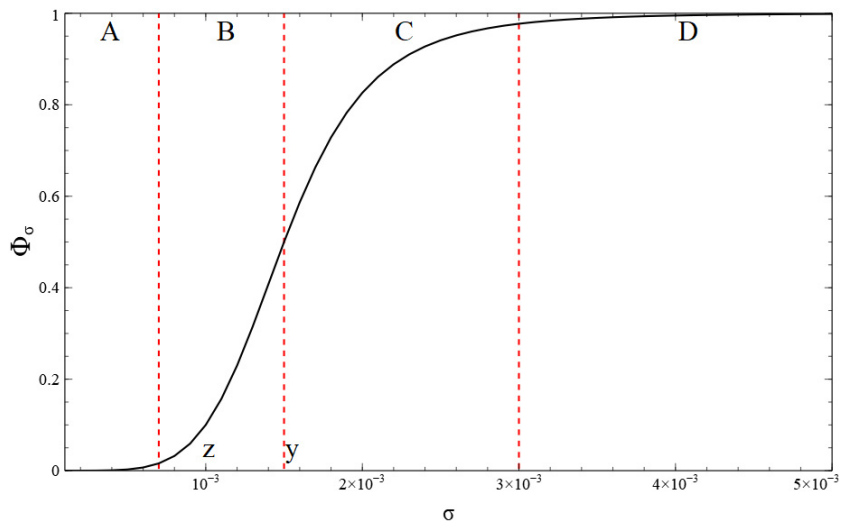

Fig. 22. (Color online) Showing the scaling of the $\sigma$ figure of merit divided into four sections. A: A good value of the figure of merit, improvements focused on other figures of merit. B: Beyond an acceptable value of the figure of merit, diminishing returns for improvements compared to other figures of merit. C: Between unacceptable and good value of figure of merit, increasing gains for improvement in the figure of merit. D: Unacceptable figure of merit.

### 4.3. Antenna figure of merit

The overall antenna figure of merit, $\Psi$, is constructed using the individual figures of merit using a weighted sum of squares:

$$\Psi = \sqrt{W_\sigma \Phi_\sigma^2 + W_\alpha \Phi_\alpha^2 + W_\beta \Phi_\beta^2 + W_\gamma \Phi_\gamma^2 + W_\delta \Phi_\delta^2}, \tag{19}$$

where $\Phi_i$ are the scaled figures of merit discussed previously and $W_i$ are selected weights for each figure of merit.

## 5. Example Case: REACH Dipole

### 5.1. Initial design types

The initial design stage required the choice of antenna style. For this a selection of antennas either already used in the 21 cm experiments, shown in Fig. 23, monopoles (Singh *et al.*, 2018b), dipoles (Bowman *et al.*, 2018), or conical log-spirals (Dyson, 1965; Sokolowski *et al.*, 2015), in addition to antennas used within other frequency bands, such as sinuous (Buck & Filipovic, 2008) and ridged horn antennas (Lee & Smith, 2004), were considered. For each of these design types, a candidate antenna was simulated with a 5 m radius circular ground plane using CST and FEKO (FEKO, 2019). The 5 m ground plane is smaller than expected to be deployed in the final instrument, however when backed by vacuum this size is large enough to correctly guide the shape of the beam while maintaining usable simulation times.

These initial designs were first roughly optimized using the previously discussed figures of merit for a global 21 cm antenna, the most promising of these where then run through a mock detection pipeline to assess the ability of the designs to perform a confident detection.

An example of the figures of merit for the antenna design styles is shown in Table 3. The monopole design was discounted due to the low inclination of beams, greater than 30° from zenith and their expected interaction with both the terrestrial environment and bright sources passing through the horizon.

Of the other two designs the conical log-spiral had better figures of merit compared to the dipole. For the dipole the possible observation range was reduced from a 4:1 to 5:2 band as the formation of multiple lobes in the beam at higher frequencies dramatically increases beam chromaticity. For the expected operational range of the dipole a detection in the mock pipeline was possible. Dipole antennas have previously been used for a detection of the global 21 cm signal, with EDGES, in addition to being geometrically simple and easy to manufacture compared to a conical log-spiral antenna.

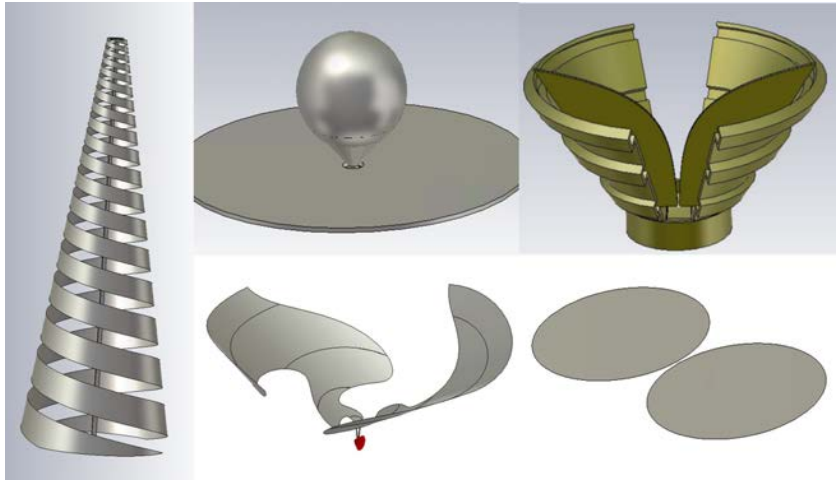

Fig. 23. (Color online) Renders of example conical log-spiral (left), monopole (top center), ridged horn (top right), inverse conical sinuous (bottom center), and dipole antennas (bottom right). For the antennas, other than the monopole where it is pictured, a 5 m circular ground plane was used for the simulations but is omitted from these images for clarity.

Table 3. A selection of the figures of merit for candidate antenna styles. Due to the preliminary stage the exact shape of $S_{11}$ is of less importance and so only the mean value and $-10\,\mathrm{dB}$ range are considered. These figures of merit do not account for the complexity of the instruments, where the conical log-spiral and horn antennas perform worse than the dipole. It should be noted that the dipole design considered here did not use a balun which would improve the reflection coefficient. The other antenna simulations include simple simulated baluns.

| Design style | Mean $S_{11}$/dB | $\beta$ | $\sigma$ |
|---|---|---|---|
| Conical log-spiral | −25 | 145 | 0.0003 |
| Wide band dipole | −12 | 90 | 0.02 |
| Horn | −15 | 150 | 0.02 |
| Monopole | −10 | 80 | 0.06 |
| Sinuous | −5 | 0 | 2 |

For these reasons the decision was made to proceed with a dipole design initially to establish and test the methodology used for the REACH experiment. Once the additional understanding has been gained from the deployment and operation of a low band dipole a conical log-spiral design will be added using the refined methods from the initial deployment.

## 5.2. *Dipole parameter selection*

The design of the dipole can be divided into several selections of parameters. The first section is to govern the size and shape of the antenna blades. The second is to describe the balun structure required for converting to a single ended feed and matching to the receiver structure. Finally the ground planes size, distance between the blades and antenna height are accounted for.

Eight parameters are used to define the seven pointed blade designs for the REACH dipole, as seen in Fig. 24. This choice was made to imitate an elliptical antenna while simplifying the manufacturing and computational process by removing the curved surfaces.

The balun structure requires a further three parameters, governing the open-ended short, radius of tubes shielding the coaxial cables, and the connecting bar between these shields. Detailed in Table 4 are all 14 parameters used to define and optimize the physical dimensions of the REACH dipole, the range of values used for these parameters is given in Table 5. For the initial sample generation a Latin hypercube is used to evenly sample over all of the parameters required to define the antenna. Parameter ranges for the antenna blade were chosen around a quarter wavelength blade at 75 MHz, as the most promising frequency for a global 21 cm signal. To identify possible interactions between ground plane dimensions and the dipole blades the radius is allowed to vary between 5 and 15 m, with the upper limit chosen to maintain reasonable simulation time. The baluns parameters are centered at the expected quarter wavelength value for the center of the frequency band.

## 5.3. *Balun design*

The REACH dipole antenna uses a Roberts style balun (Fig. 25) (Roberts, 1957) similar to that used an open short of the other blade via a printed circuit board bridge. The length of this open stub has the largest tuneable impact upon the impedance transformation of the balun.

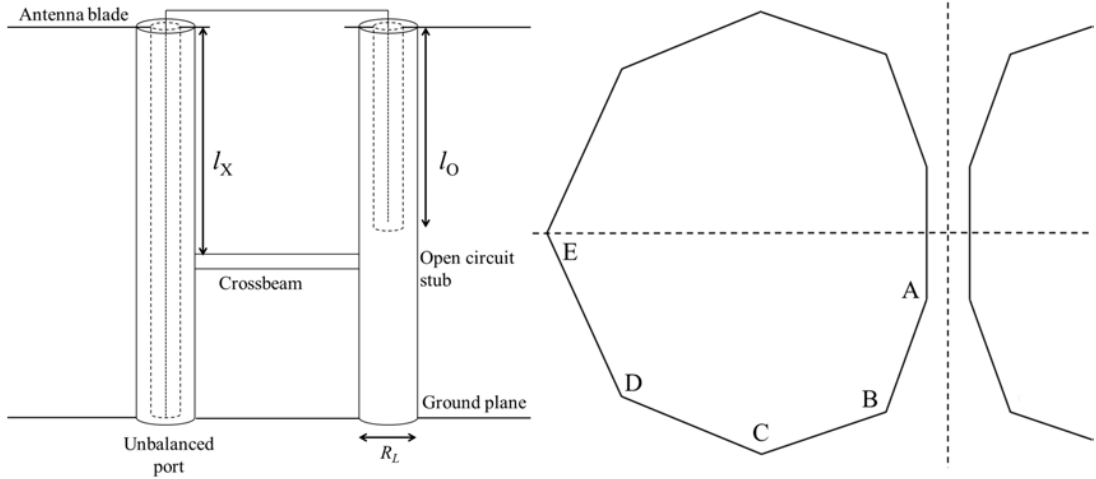

Fig. 24.  Left: The design of the Roberts balun used for the REACH dipole. Using a coaxial cable connection to the receiver through one of two shielded legs. The second leg, in addition to providing symmetry, contains an open ended coaxial cable used to refine the balun impedance transformation. Right: The generic seven pointed shape used to describe the shape of the dipole, designed to emulate an elliptical dipole while removing the need for curved edges which increase simulation complexity and time.

Table 4.  The parameters used for the design of the REACH dipole.

| Name | Antenna section |
|---|---|
| $A_x, B_x, B_y, C_x, C_y, D_x, D_y, E_x$ | Blade corners |
| $l_X, l_O, R_L$ | Balun and leg structure |
| $R_G$ | Circular ground plane size |
| $h$ | Blade height |
| Gap | Blade separation |

Table 5.  The parameter ranges for the initial hypercube sampling for figure of merit analysis used for the REACH dipole.

| Name | Minimum/mm | Maximum/mm |
|---|---|---|
| $A_x$ | 16 | 19 |
| $B_x$ | 68 | 85 |
| $B_y$ | 565 | 640 |
| $C_x$ | 265 | 460 |
| $C_y$ | 495 | 595 |
| $D_x$ | 620 | 780 |
| $D_y$ | 445 | 550 |
| $E_x$ | 850 | 920 |
| $l_X$ | 415 | 479 |
| $l_O$ | 430 | 440 |
| $R_L$ | 22 | 31 |
| $R_G$ | 5000 | 15,000 |
| $h$ | 650 | 900 |
| Gap | 11 | 14 |

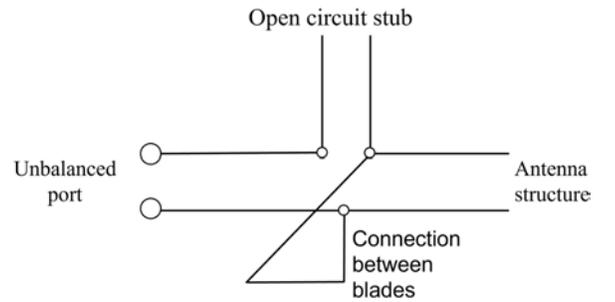

Fig. 25.  A circuit diagram for the Roberts balun, showing the main contributions to the impedance transformation, the open short and parallel transmission line formed within the leg structure.

The balun parameters are optimized during the same sequence as the antenna blades, to allow as much synergy as possible between the sections of the design. A resonant dipole presents an absolute impedance of $73\,\Omega$ which is frequency-dependent. The addition of a balun to the dipole structure allows the balun to act as a transformer and may be tuned to better match the antenna impedance to the output $50\,\Omega$ cable.

### 5.4.  *REACH figures of merit*

For the REACH antenna five figures of merit are used for the design process, detailed in Table 6.

in the EDGES2 antenna. The schematic of this balun is shown in Fig. 24, comprising of a coaxial feed from which the outer section is connected to the antenna blade and center wire is then connected to

Table 6. The five numerical figures of merit used for the design of the REACH antenna, with the smoothness of the antenna beam and the matching of the antenna impedance to a standard $50\,\Omega$ receiver. The goal values listed here are for the first iteration and can be expected to be modified as more is learned about the requirements for a successful detection over the course of the REACH experiment.

| Figure of merit | Defining equation | Design aspect | Goal value |
|---|---|---|---|
| $\sigma$ | (16) | Smoothness of antenna beam | Minimize to 0 |
| $\alpha$ | (11) | Mean value of reflection caused by antenna impedance | Minimize to $-10\,$dB |
| $\beta$ | (12) | Range of values with an acceptable $S_{11}$ value | Maximize to 90 MHz |
| $\gamma$ | (13) | Flatness and depth of reflection coefficient | Minimize to 0% |
| $\delta$ | (15) | Minimum frequency for $-10\,$dB $S_{11}$ | Minimize to 50 MHz |

Table 7. The preferential values, for calculation of $\Phi_i$ used for scaling the figures of merit for the REACH antenna.

| Figure of merit | $y$ | $z$ |
|---|---|---|
| $\sigma$ | 0.0015 | 0.001 |
| $\alpha$ | $-10$ | $-12$ |
| $\beta$ | 50 | 70 |
| $\gamma$ | 50 | 30 |
| $\delta$ | 60 | 55 |

Table 7 gives the values of $y$ and $z$ used for scaling the respective figures of merit for combination into the overall figure of merit $\Psi$. The goal values shown here were selected using results achieved from the initial antenna design exploration, representing the most promising results achieved during this process. It is planned for these goal values to be updated as more is learned about the REACH experiment and the requirements for a confident detection. To assist with the initial deployments more stringent and time-consuming tests, for example a mock detection pipeline, were used once promising candidate antennas were identified. Due to diminishing returns or physical limitations the "optimal value" of some of the figures of merit is not the same as the value taken for $y$, for example $\sigma$ where there is expected to be some chromaticity present in the beam. As the analysis pipeline is able to compensate for a small level of chromaticity it becomes preferable to optimize other figures of merit. Figure 22 shows the scaling function for $\sigma$ divided into four rough regions demonstrating good, acceptable, improving, and unacceptable figures of merit.

For the design of each antenna the priority of the previously described figure of merits will vary depending on the intended use case. The weighting of the overall figure of merit and choice of scaling functions allow for tuning of the prioritized figures of merit. The weightings allow for balancing of aspects of the antenna within the figure of merit, for instance the initial weighting of $\Psi$ used for the REACH antenna is $[W_\sigma, W_\alpha, W_\beta, W_\gamma, W_\delta] = [4, 1, 1, 1, 1]$, with a weight for each of the previously discussed figures of merit. This was chosen initially to balance the contributions from the beam chromaticity and the antenna impedance reflections. As REACH is deployed and begins to take measurements these weightings can be tuned to emphasize the optimization of different aspects of the antenna. If beam chromaticity is the critical barrier to a detection increasing $W_\sigma$ can be used to generate a design with a superior chromaticity figure of merit, at the expense of more difficult calibration; and similarly for the other figures of merit. As with the goal values in Table 6 the weightings discussed in this paper are chosen as a starting point and expected to be developed over time.

To visualize $\Psi$ corner plots such as Fig. 26 are used. These plots show the value of the figure of merit function evaluated over grids between pairs of physical parameters. For this figure the mean value of the other 12 parameters is used. This allows for an understanding to be gained of possible interactions between the many parameters of the antenna design, with lower values of figure of merit preferable.

Using the weightings it is informative to consider the two separate parts of the figure of merit, shown in Fig. 27 using $W = [1, 0, 0, 0, 0]$ and $W = [0, 1, 1, 1, 1]$ to consider only the beam chromaticity and the overall impedance of the antenna. The weighting $W = [0, 0, 0, 0, 1]$ is also highly informative indicating the key components influencing lowest antenna operating frequency.

Figure 26 starts to give an idea of the relationship between physical parameters and figures of merit. This information can be used for more targeted refinement of the antenna in the field and for later iterations as further limitations for a detection are identified.

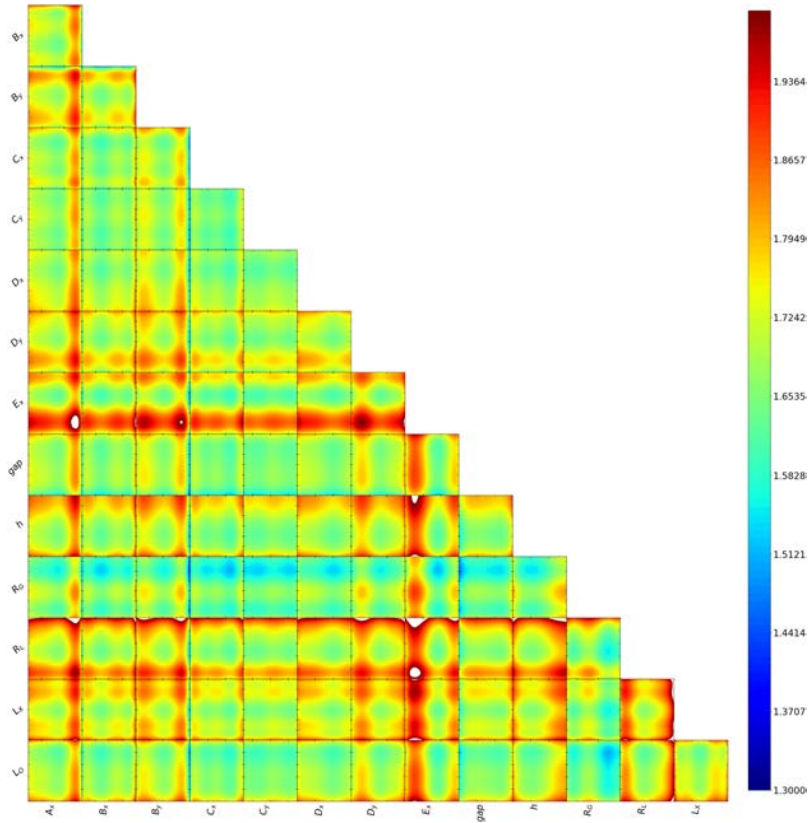

Fig. 26. (Color online) Full figure of merit compared to parameters, using $W = [4, 1, 1, 1, 1]$, for this plot the parameters not relevant to each subplot are set to their mean value. Here, the individual figure of merit values generated for each antenna of the initial hypercube have been fitted using a sixth-order polynomial over the physical parameters. For this plot each square is a slice through the mean value of the other parameters in the hyperspace. This plot allows the identification of promising regions within the parameter space, shown in blue. Using numerical methods a minimum point of these polynomials is found as the expected best parameter set for the given constraints. The overall figure of merit is shown ranging from blue to red.

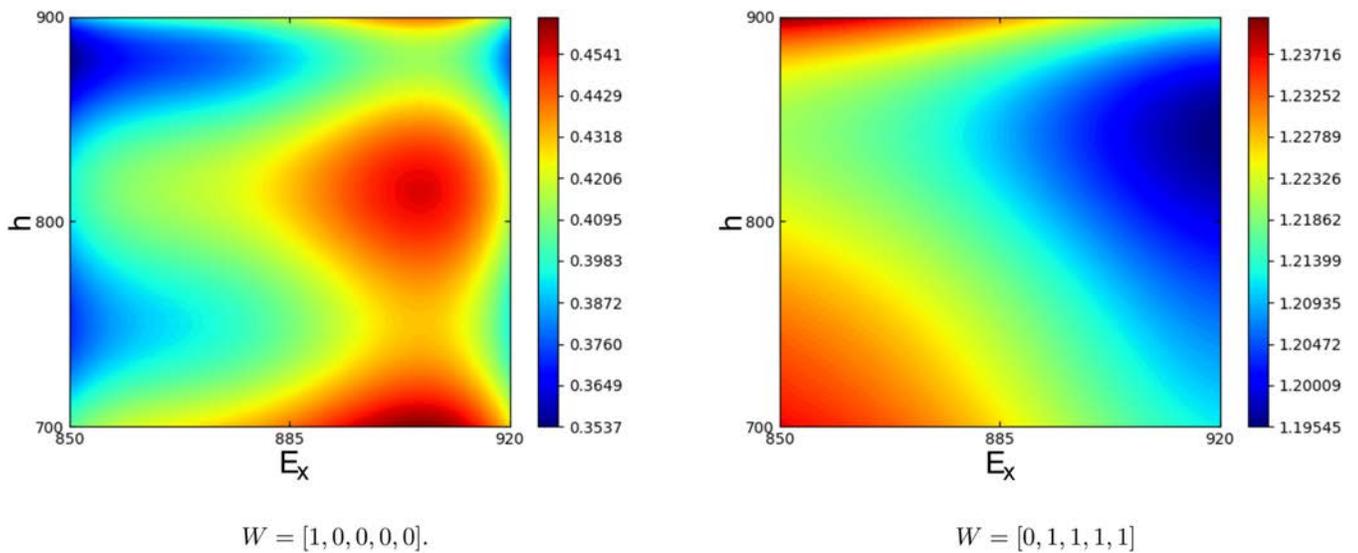

$W = [1, 0, 0, 0, 0]$.          $W = [0, 1, 1, 1, 1]$

Fig. 27. (Color online) Two slices of figure of merit comparing the chromaticity (left) and combined $S_{11}$ (right) weightings. Both slices are taken with all other parameter values at their mean for the full parameter range. A demonstration of the differing preferences of the figure of merit, with an increased Ex, and height assisting $S_{11}$ whereas chromaticity benefits from a lower height and middling Ex around 850 mm.

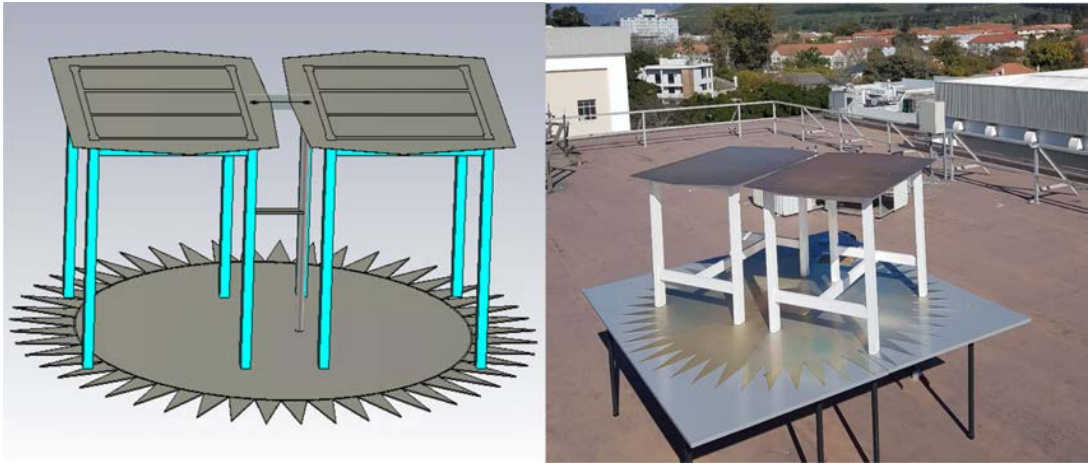

Fig. 28.   (Color online) Left: The CST simulation model of the prototype antenna, used for verification of the simulations used for the design of the dipole antenna. Right: A prototype antenna constructed at Stellenbosch University, prior to the addition of a balun. Due to materials and space available a smaller ground plane was used than in standard simulations or the final deployment.

Seen in Fig. 27 is an example of one trade off required to be made for a global 21 cm antenna. Here, it is seen that the reduced antenna height is favored by the chromaticity figure of merit, but unfavorable by the combined $S_{11}$ figures of merit. These plots also demonstrate the possibility of regional preference for the figures of merit. When considering these plots it is important to recall that they are 2D cut through slices of a 14-dimensional space, parametrized by the physical parameters of the antenna.

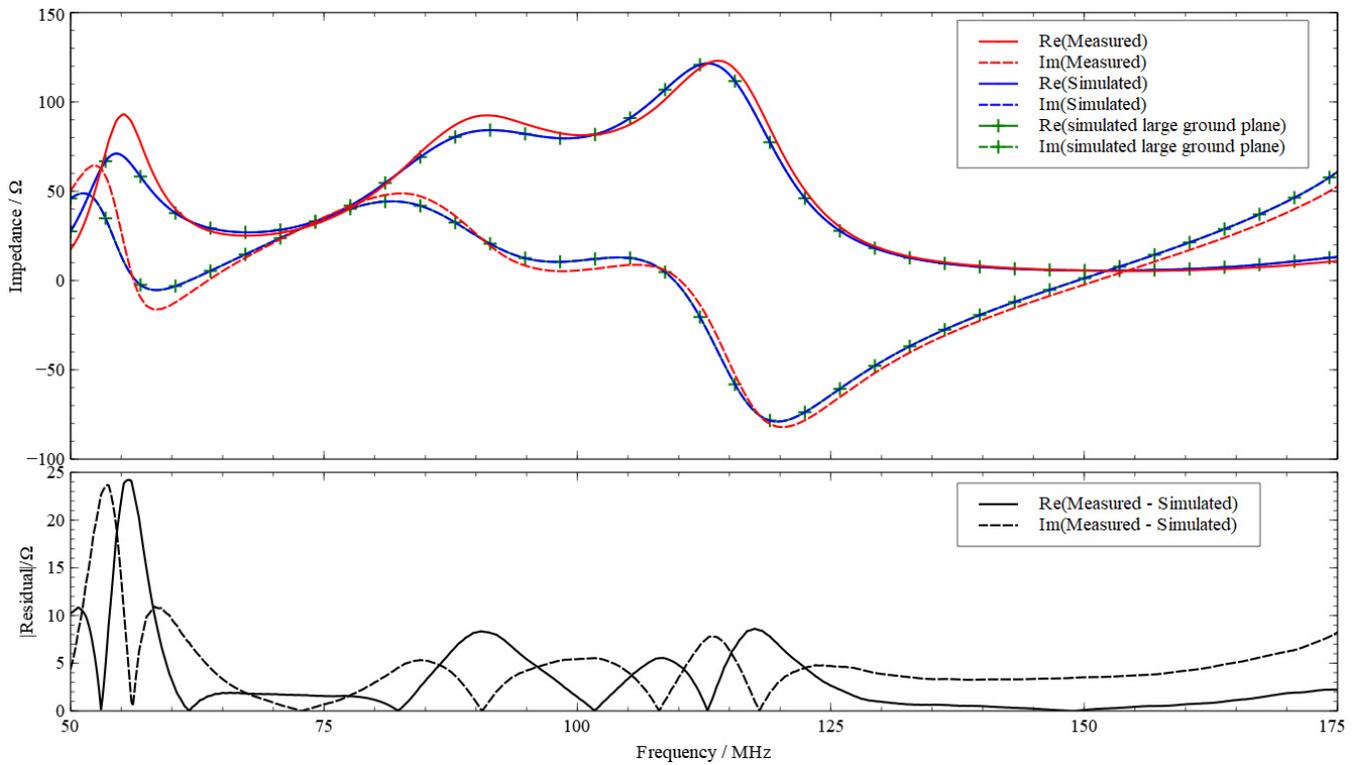

Fig. 29.   (Color online) Top: The impedance of a prototype antenna, compared to simulations of the same antenna with a 2 m circular serrated ground plane and a 10 m smooth circular ground plane. There is negligible difference between the simulations using different sizes for ground planes. There is some difference seen at the peaks of resistance for the measured antenna, at 55 and 90 MHz, however overall the simulations and measurement are sufficiently close. Bottom: The residual difference between simulated and measured measured prototype antennas, two spikes are seen around 55 MHz to 25 Ω discrepancy, and otherwise the residual seen is within 10 Ω. This is considered sufficiently close for the simulations to be considered accurate for the process of designing an antenna.

A functional minimization is used to find the minima of the polynomial fitted to the figure of merit to identify the most promising candidate which is then to be evaluated in more detail using prototypes and a mock detection pipeline.

### 5.5. *Prototype*

For the REACH experiment calibration of the antenna impedance will be done through dynamic on site measurement (Roque *et al.*, 2021). However for verification of the computational methods in addition to manufacturing and construction procedures to be established, a preliminary antenna design was constructed at Stellenbosch University South Africa, shown in Fig. 28. The reflection coefficient of this prototype was measured using a 910 mm radius circular ground plane with 48,300 mm triangular serrations. For this measurement an extra length of cable was required below the antenna. For comparison to simulations an ideal coaxial transmission line of length 240 mm was added in series after the antenna for calculation of $S_{11}$.

The ground plane used for the measurement was smaller than the expected final ground plane. Ground plane size is expected to have negligible effect upon the antenna impedance. To verify this the antenna was simulated with a 10 m circular ground plane for additional comparison to the smaller ground plane simulation and measurement. As shown in Fig. 29, the difference between the two simulations is smaller than the difference between either simulation and the measurement.

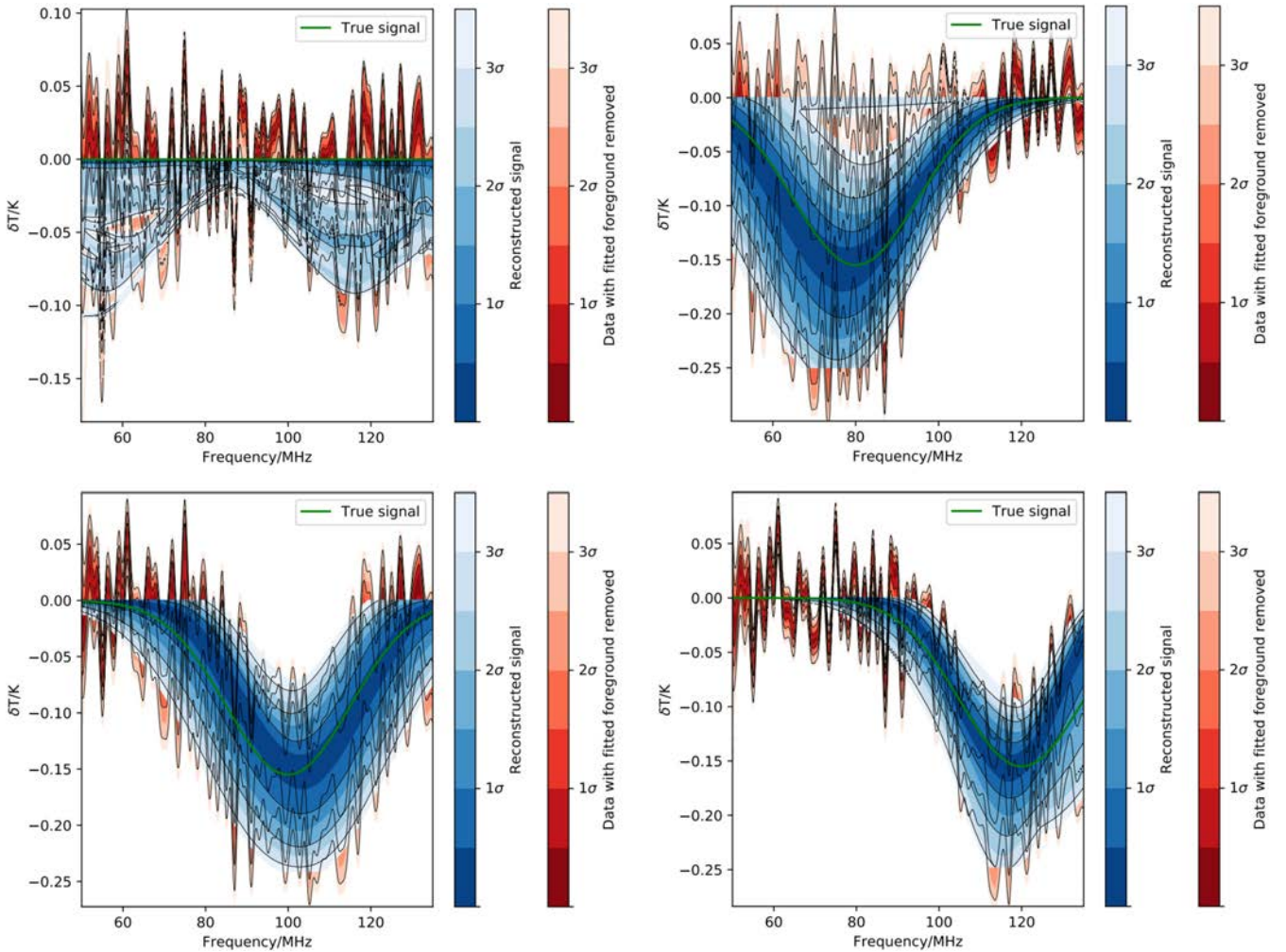

Fig. 30. (Color online) Attempted reconstructions of four different injected signals using the prototype antenna. Top left: Null signal showing no false detection is provided by the antenna (2.5:1 odds confidence to reject a signal). Top right, bottom left, bottom right: Three injected Gaussian signals, green line, reconstructed, blue, the variance seen in the reconstructed signal is expected to be reduced with increased frequency resolution and observation time.

Figure 29 shows the impedance for the prototype antenna. This figure shows that there is very close agreement between the two simulated antennas of varying ground plane sizes and the measured prototype. The difference in impedance for the simulations including different ground planes is negligible. Here, it is seen that there is some deviation with a lag of 10 MHz between the real and imaginary parts of the impedance, with the largest deviation occurring at 65 MHz. This deviation is small enough to continue using the computational models for the antenna for suggestions to the design. In the field an active calibration will be used to measure the reflection coefficient for the antenna, so general features of the simulated reflection are of more import than the exact values.

### 5.6. *Mock detection pipeline results*

To assess the ability of this style of antenna to perform a possible detection of the global 21 cm signal a CST model of the prototype antenna, Fig. 28, was placed through a mock analysis pipeline. Although this simulation uses only the small ground plane and a vacuum backing the addition of these features to a final calculation are expected to improve the ability of the antenna to perform a possible detection.

This analysis was calculated using the same method as detailed in Anstey *et al.* (2021a, 2021b). This method attempts to detect an injected Gaussian signal using Bayesian analysis to assess the parameters of candidate Gaussian functions being present within the data. This method replicates the method of searching for candidate 21 cm signals to be used in the final analysis. In Fig. 30 are shown the returned fit using no injected signal and three injected Gaussian signals with standard deviation 15 MHz, depth 155 mK and center frequencies 80, 100, and 120 MHz. The sky map was taken from 00:00 to 01:00 on January 1st, 2019 as viewed from the Karoo Astronomy Reserve and the frequency range observed over 50–135 MHz resolution of 1 MHz. With real observations the frequency resolution and observation time will be increased, which is expected to reduce the uncertainty seen in Fig. 30. Using the sample Gaussian signal the antenna was able to detect all center frequencies with confidence and with a 2.5:1 odds confidence reject a signal when there is none present. This ability of the antenna to confidently detect signals across the operational frequency range with both the data analysis and physical construction of the antenna still to be improved is a good verification of the antennas capability.

## 6. Conclusions

This paper considered the general characteristics of candidate antennas for a global 21 cm experiment noting benefits and trade-offs for the design styles. Showing the importance of coupling to the sky model used and effect of beamwidth on the accuracy of this measurement. Following this the figures of merit required for an effective global 21 cm antenna were discussed.

Also was detailed a quantitative, physically based, figure of merit approach for the design of a global 21 cm antenna. Starting with multiple design styles before selecting and refining a dipole design. The current prototype design shows a 2.5:1 observation band from 55 MHz and antenna gain patterns able to detect a range of injected signals across the 80–130 MHz center frequency at 155 mK amplitude and correctly identify a null signal. A deployment is expected to take place in the Karoo Radio Astronomy Reserve South Africa during 2021, with observations beginning shortly afterwards.

### Acknowledgments

The authors would like to thank the Kavli Foundation for their support of REACH. John Cumner was supported by EPSRC iCASE in partnership with BT. Eloy de Lera Acedo was supported by the Science and Technologies Facilities Council. Dirk de Villiers was supported in part by the South African Radio Astronomy Observatory, which is a facility of the National Research Foundation, an agency of the Department of Science and Technology (Grant Number 75322). The work of Girish Kulkarni and Shikhar Mittal was supported by the Department of Atomic Energy (Government of India) research project under Project Identification Number RTI 4002 and also in part by the Max Planck Society via a partner group grant.